\documentclass[a4paper,11pt]{article}
\pdfoutput=1 

\usepackage{jcappub} 

\usepackage[T1]{fontenc} 
\usepackage{ifthen}
\usepackage{epsfig}
\usepackage{booktabs} 
\usepackage{natbib}
\usepackage{bbold}
\usepackage{diagbox}
\usepackage{slashed}

\usepackage{feynmp-auto}
\usepackage{subfig}

\usepackage{capt-of}

\renewcommand{\toprule}{\specialrule{1.5pt}{0em}{1pt} \midrule}
\renewcommand{\bottomrule}{\midrule \specialrule{1.5pt}{1pt}{0em}}

\usepackage{floatrow}
\newfloatcommand{capbtabbox}{table}[][\FBwidth]

\newcommand{\ii}{i}

\newcommand{\dd}{\mathrm{d}}

\renewcommand{\epsilon}{\varepsilon}

\renewcommand{\vec}[1]{\boldsymbol{#1}}
\newcommand{\boldell}{\boldsymbol{\ell}}
\renewcommand{\bar}{\overline}
\renewcommand{\Re}{\mathrm{Re}}
\renewcommand{\Im}{\mathrm{Im}}

\newcommand{\vt}{\vec{v}^{\bot}}

\newcolumntype{L}{>{$}l<{$}}
\newcolumntype{C}{>{$}c<{$}}

\newcommand{\ket}[1]{\left|#1\right\rangle}

\newcommand{\vPerpEl}{\vec v_{\rm el}^\perp}
\newcommand{\vvp}{\vec v_{\rm el}^\perp}
\newcommand{\frqm}{\frac{\vec q}{m_e}}
\newcommand{\qAbs}{|\vec q|}
\newcommand{\vAbs}{|\vec v_{\rm el}^\perp|}
\newcommand{\vA}{ \Re \left( \boldsymbol{f}^{*}_{1\rightarrow 2}f_{1\rightarrow 2} \right) }
\newcommand{\cvff}{\boldsymbol{f}_{1\rightarrow 2}}
\newcommand{\fAbs}{|\boldsymbol{f}_{1\rightarrow 2}|}
\newcommand{\NRop}[1]{\langle\Hat{\mathcal{O}}_{#1}\rangle} 
\newcommand{\ubar}{\bar{u}}
\newcommand{\rd}{\ensuremath{\,\mathrm{d}}}
\allowdisplaybreaks

\usepackage{ccicons}
\usepackage{listings}
\usepackage{color}   
\usepackage{braket} 
\usepackage{hyperref}
\usepackage{dsfont}  
\usepackage{leftidx} 
\RequirePackage{mathrsfs}
\RequirePackage{dsfont}
\usepackage{bm}
\usepackage{placeins}
\usepackage{float}

\title{\boldmath Dark matter - electron interactions in materials beyond the dark photon model}

\author[a]{Riccardo Catena}
\author[a]{Daniel Cole}
\author[b]{Timon Emken}
\author[c]{Marek Matas}
\author[c]{Nicola Spaldin}
\author[c]{Walter Tarantino}
\author[a]{and Einar Urdshals}

\affiliation[a]{Chalmers University of Technology, Department of Physics, SE-412 96 G\"oteborg, Sweden}
\affiliation[b]{The Oskar Klein Centre, Department of Physics, Stockholm University, AlbaNova, SE-10691 Stockholm, Sweden}
\affiliation[c]{Department of Materials, ETH Z\"urich, CH-8093 Z\"urich, Switzerland}

\emailAdd{catena@chalmers.se}
\emailAdd{timon.emken@fysik.su.se}
\emailAdd{marek.matas@mat.ethz.ch}
\emailAdd{nicola.spaldin@mat.ethz.ch}
\emailAdd{wtarantino@dsf.unica.it}
\emailAdd{urdshals@chalmers.se}

\abstract{The search for sub-GeV dark matter (DM) particles via electronic transitions in underground detectors attracted much theoretical and experimental interest in the past few years.~A still open question in this field is whether experimental results can in general be interpreted in a framework where the response of detector materials to an external DM probe is described by a single ionisation or crystal form factor, as expected for the so-called dark photon model.~Here, ionisation and crystal form factors are examples of material response functions:~interaction-specific integrals of the initial and final state electron wave functions.~In this work, we address this question through a systematic classification of the material response functions induced by a wide range of models for spin-0, spin-1/2 and spin-1 DM.~We find several examples for which an accurate description of the electronic transition rate at DM direct detection experiments requires material response functions that go beyond those expected for the dark photon model.~This concretely illustrates the limitations of a framework that is entirely based on the standard ionisation and crystal form factors, and points towards the need for the general response-function-based formalism we pushed forward recently~\cite{Catena:2019gfa,Catena:2021qsr}.~For the models that require non-standard atomic and crystal response functions, we use the response functions of~\cite{Catena:2019gfa,Catena:2021qsr} to calculate the DM-induced electronic transition rate in atomic and crystal detectors, and to present 90\% confidence level exclusion limits on the strength of the DM-electron interaction from the null results reported by XENON10, XENON1T, EDELWEISS and SENSEI.} 

\begin{document} 
\maketitle
\flushbottom

\section{Introduction}

The nature of the dark matter (DM) component of the universe, whose gravitational effects are visible from local to cosmological scales, remains elusive.~The leading hypothesis in astroparticle physics is that DM is made of new, yet undetected particles~\cite{appec}.~Direct detection experiments play a crucial role in the quest for the DM particle, and search for DM scattering events in low-background detectors located deep underground~\cite{Drukier:1983gj,Goodman:1984dc,Undagoitia:2015gya}.~So far, this class of experiments has focused on the search for the Weakly Interacting Massive Particle (WIMP):~a DM candidate of mass between few GeV and tens of TeV that interacts with the known particles with weak scale interactions, and predicted by theories which were not initially formulated in a DM context, e.g.~supersymmetry~\cite{Arcadi:2017kky,Roszkowski:2017nbc}.~The lack of an unambiguous WIMP detection has recently motivated a systematic, theoretical and experimental exploration of alternative paradigms~\cite{Mitridate:2022tnv}.

An alternative scenario that could explain why WIMPs have so far escaped detection is the one where DM is lighter than the nucleons bound to nuclei, and thus too light to induce an observable nuclear recoil in a direct detection experiment~\cite{Battaglieri:2017aum}.~An interesting aspect of this scenario is that it is testable, as a Milky Way DM particle in the keV-GeV mass range would have enough kinetic energy to cause an observable electronic transition in a detector material, provided it interacts strongly enough with the constituent electrons~\cite{Essig:2011nj}.~Being a testable explanation for the lack of WIMP discovery, the sub-GeV DM paradigm has recently attracted much attention~\cite{Kahn:2021ttr,Mitridate:2022tnv}.~In particular, a variety of experimental tests has been proposed or performed to probe this idea.~This includes the search for: atomic ionisations in noble gas xenon and argon detectors~\cite{Essig:2011nj,Essig:2017kqs,DarkSide:2018ppu,Catena:2019gfa,Aprile:2019xxb,XENON:2020rca}, electronic transitions in semiconductor crystals~\cite{Graham:2012su,Essig:2015cda,Derenzo:2016fse,Agnese:2018col,Kurinsky:2019pgb,DAMIC:2019dcn,EDELWEISS:2020fxc,SENSEI:2020dpa,Griffin:2020lgd,Catena:2021qsr,Griffin:2021znd,Knapen:2021run,Hochberg:2021pkt,Lasenby:2021wsc,Chen:2022pyd} as well as in superconductors~\cite{Hochberg:2015pha,Hochberg:2021yud} and 3D Dirac materials~\cite{Hochberg:2017wce,Geilhufe:2019ndy,Coskuner2021Jan}, electron ejections from graphene~\cite{Hochberg:2016ntt} and carbon nanotubes~\cite{Cavoto:2019flp}, and collective phenomena such as phonons~\cite{Knapen:2017ekk,Trickle:2019nya} and magnons~\cite{Trickle:2019ovy}.~This is an incomplete list, and we refer to~\cite{Kahn:2021ttr,Mitridate:2022tnv} for an extended review of the field.

A central element in the development and implementation of the experimental tests mentioned above is the theoretical modelling of DM-electron interactions.~Typically, the interactions between DM particles and electrons are described within the ``dark photon model'', which extends the Standard Model of particle physics by an additional $U(1)$ gauge group under which only the DM particle is charged~\cite{Holdom:1985ag,Battaglieri:2017aum}.~In this model, the DM-electron interaction arises from the ``kinetic mixing'' between the ordinary photon and the gauge boson associated with the new $U(1)$ group, i.e.~the dark photon.~A crucial prediction of the model is that the response of materials to DM-electron interactions depends on a single, material-specific ionisation or crystal form factor~\cite{Essig:2015cda}.~While the dark photon model has extensively been used in the interpretation of results from DM direct detection experiments, it is also a rather restrictive framework, as it a priori neglects DM-electron couplings that are allowed by observations, and assumes that the interactions between DM and electrons are necessarily mediated by a spin-1 particle.~Furthermore, the model is often complemented by the additional assumption that the DM particle has either spin 0 or 1/2.~However, there is no experimental evidence supporting these restrictions and, therefore, a more general approach is in many respects preferable.

In order to develop a description of DM-electron scattering in detector materials that goes beyond the one of the dark photon model, we recently developed a non-relativistic effective theory for the scattering of Milky Way DM particles by the electrons bound to isolated atoms~\cite{Catena:2019gfa} or semiconductor crystals~\cite{Catena:2021qsr}.~This allowed us to systematically classify the interactions between DM particles and electrons in terms of a finite set of Galilean and rotational invariant quantum mechanical operators that are defined in the DM-electron spin space.~Virtually any DM particle theory can be mapped onto a linear combination of such operators with momentum transfer dependent coefficients.~Importantly, this general approach to the theoretical modelling of DM-electron interactions predicts that up to four (for isolated atoms) and up to five (for crystals) electron wave function overlap integrals, or ``response functions'', are required in order to accurately describe the scattering of DM particles by the electrons bound in detector materials.~In other words, the response of materials to an external DM perturbation can in principle be significantly different from the one predicted by the dark photon model.~Recently, our formalism was adopted by the XENON collaboration to analyse their latest S2-only data from XENON1T~\cite{XENONCollaborationSS:2021sgk}.

A still open question in this field is whether the novel material response functions we identified in~\cite{Catena:2019gfa} and~\cite{Catena:2021qsr} are indeed  important in the interpretation of DM direct detection experiments, or they only appear in marginal models.~While in~\cite{Catena:2019gfa,Catena:2021qsr} we provided relevant examples, such as the anapole or magnetic dipole DM models, where our ``response function'' formalism is crucial to accurately predict observable electronic transition rates, a systematic classification of the DM-electron interaction models that can generate a non-standard material response to an external DM perturbation is still missing.~Answering this question is important, as it would clarify whether present and future DM direct detection experiments can indeed be interpreted within a framework that entirely relies on a single ionisation or crystal form factor, or a general analysis of the response to DM-electron interactions of detector materials is crucial.

In this work, we answer the above question by calculating the rate of electronic transitions induced by the non-relativistic scattering of DM particles by electrons bounds to isolated atoms or silicon and germanium crystals for a wide range of models for spin-0, spin-1/2 and spin-1 DM.~Remarkably, we find eight relativistic Lagrangians which in the non-relativistic limit predict a non-standard material response to external DM-electron interactions.~This corroborates the results we reported in~\cite{Catena:2019gfa,Catena:2021qsr} and shows the importance of a ``response function'' formalism in the interpretation of DM direct detection experiments.~This work is organised as follows.~In Sec.~\ref{sec:lagrangians}, we introduce the relativistic models for DM-electron interactions we are interested in.~In Sec.~\ref{sec:dd} we  calculate the rate of DM-induced electronic transitions in atoms or crystals predicted by the models of Sec.~\ref{sec:lagrangians} within the response function formalism of~\cite{Catena:2019gfa,Catena:2021qsr}.~Here, we also show that the response functions for xenon, germanium and silicon can be expressed within a unified formalism that allows for a direct comparison of different target materials.~Finally, in Sec.~\ref{sec:results}, we compare our theoretical predictions with the null results reported by the XENON10~\cite{Angle:2011th}, XENON1T\cite{Aprile:2019xxb}, EDELWEISS~\cite{EDELWEISS:2020fxc} and SENSEI~\cite{SENSEI:2020dpa} DM direct detection experiments, setting 90\% confidence level upper bounds on the coupling constants of the models we consider in this work.~We summarise and conclude in Sec.~\ref{sec:conclusion}, and refer to three appendices for the details of our calculations, useful identities and complementary results.

\section{General models for dark matter-electron interactions}
\label{sec:lagrangians}
This section introduces the framework we use to describe DM-electron interactions in detector materials.

In the interpretation of data from direct detection experiments, DM-electron interactions are typically described within the so-called dark photon model~\cite{Holdom:1985ag} (for a review, see also~\cite{Battaglieri:2017aum} and references therein).~This framework extends the Standard Model of particle physics by an additional $U(1)$ gauge group and by a DM candidate.~Only the DM particle is charged under the new gauge group, while a ``kinetic mixing'' between ordinary and dark photon generates the interactions between the electrically charged particles in the Standard Model and the DM candidate, which, within this framework, typically consists of either scalar or fermionic particles.~Here, by ``dark photon'' one refers to the gauge boson associated with the new $U(1)$ group.

In this work, we extend the standard dark photon model in three ways.~First, we consider both spin-0 and spin-1 particles as ``mediators'' of the DM-electron interaction, without restricting ourselves to vector mediators, as in the dark photon model.~Second, we focus on DM-mediator interaction vertices with a general Lorentz structure.~In contrast, in the dark photon model only interactions arising from a kinetic mixing are considered.~Finally, we allow for DM to have spin 0, spin 1/2 or spin 1.~These extensions are motivated by the lack of an experimentally preferred model for sub-GeV DM, which, in our view, should be compensated by a general approach to the theoretical modelling of DM-electron interactions.~The main implication of these extensions is that in the models we consider here the response of detector materials to DM-electron interactions cannot be described via the canonical ``form factors'' introduced within the dark photon model.~As we will see below, it must be modelled within the ``response function'' formalism we developed by using effective theory methods in~\cite{Catena:2019gfa}, for isolated atoms, and in~\cite{Catena:2021qsr}, for crystals.~From this point of view, the analysis presented here provide a crucial, concrete application of the formalism we established in~\cite{Catena:2019gfa,Catena:2021qsr}, which is therefore proven to be relevant and important in the theoretical modelling of DM-electron interactions.~Furthermore, it also shows how the effective operators identified in~\cite{Catena:2019gfa,Catena:2021qsr} can arise from the non-relativistic reduction of relativistic theories for DM-electron interactions.~Below, we introduce the models we are interested in, focusing on scalar, fermionic and vector DM separately.

Notice that the models we consider here do not provide a mechanism for the DM particle and mediator mass generation, and should therefore be completed in the high-energy (i.e. ultra-violet) limit.~While this completion is expected to introduce new, phenomenologically interesting particles, it goes beyond the purposes of this work.~Therefore, we refer to~\cite{Arcadi:2017kky} and references therein for a review of possible ultra-violet completions of the models introduced below. 

\subsection{Scalar dark matter}
\label{sec:scalar}
Let us start by introducing our models for spin-0 DM.~Here, we denote by $S$ the complex scalar that describes the DM particle, and consider two cases for the particle that mediates the DM-electron interactions, i.e.~``the mediator''.~In the first case, we assume a spin-0 particle of mass $m_\phi$ described by the real scalar field $\phi$.~In the second case we assume a spin-1 particle of mass $m_G$ described by the real vector $G_\mu$.~In the case of a scalar mediator, $\phi$, we introduce the Lagrangian 
\begin{eqnarray}
\mathcal{L}_{S\phi e} &=& \partial_\mu S^\dagger\partial^\mu S - m_S^2S^\dagger S - \frac{\lambda_S}{2}(S^\dagger S)^2 \nonumber\\
&+&\frac{1}{2}\partial_\mu\phi\partial^\mu\phi - \frac{1}{2}m_\phi^2\phi^2 -\frac{m_\phi\mu_1}{3}\phi^3-\frac{\mu_2}{4}\phi^4 \nonumber\\
&+& \ii \bar{e}\slashed{D} e - m_e \bar e e \nonumber\\
&-&g_1m_SS^\dagger S\phi -\frac{g_2}{2}S^\dagger S\phi^2-h_1\bar e e\phi-ih_2\bar{e}\gamma^5e\phi\,, 
\label{eq:Lss}
\end{eqnarray}
where $m_S$ is the DM mass, while $\lambda_S$, $\mu_1$, $\mu_2$, $g_1$, $g_2$, $h_1$, and $h_2$ are coupling constants.~In Eq.~(\ref{eq:Lss}), $m_e$ and $e$ are the electron mass and Dirac spinor, respectively.~On the other hand, when the interaction between DM particles and electrons is mediated by a spin-1 particle, we assume 
\begin{eqnarray}
\mathcal{L}_{SGe} &=& \partial_{\mu}S^{\dagger}\partial^{\mu}S -m_S^2 S^{\dagger}S -\frac{\lambda_S}{2}(S^{\dagger}{S})^2  \nonumber \\
&-&\frac{1}{4}\mathcal{G}_{\mu\nu}\mathcal{G}^{\mu\nu} + \frac{1}{2}m_G^2G_{\mu}G^{\mu} -\frac{\lambda_G}{4}(G_{\mu}G^{\mu})^2 \nonumber \\
&+&i\bar{e}\slashed{D}e -m_e\bar{e}e  \nonumber\\
&-&\frac{g_3}{2}S^{\dagger}SG_{\mu}G^{\mu} -ig_4(S^{\dagger}\partial_{\mu}S-\partial_{\mu}S^{\dagger}S)G^{\mu} \nonumber\\
&-&h_3(\bar{e}\gamma_{\mu}e	)G^{\mu}-h_4(\bar{e}\gamma_{\mu}\gamma^5e)G^{\mu}\,,
\label{eq:Lsv}
\end{eqnarray}
where $\lambda_G$, $g_1$, $g_2$, $h_1$, and $h_2$ are coupling constants, while $\mathcal{G}_{\mu\nu}$ is the field strength tensor for $G_\mu$.

\subsection{Fermionic dark matter}
\label{sec:fermion}
We now focus on the case of spin-1/2 DM with mass $m_\chi$.~When the interactions between the DM particles and electrons are mediated by a scalar particle of mass $m_\phi$, we assume the Lagrangian,
\begin{eqnarray}
\mathcal{L}_{\chi\phi e} &=& \ii\bar{\chi}\slashed{D}\chi - m_{\chi}\bar{\chi}\chi \nonumber\\
&+&\frac{1}{2}\partial_\mu\phi\partial^\mu\phi - \frac{1}{2}m_\phi^2\phi^2 -\frac{m_\phi\mu_1}{3}\phi^3-\frac{\mu_2}{4}\phi^4 \nonumber\\
&+& \ii\bar{e}\slashed{D} e - m_e \bar e e\nonumber\\	
&-&\lambda_1\phi\bar{\chi}\chi -i\lambda_2\phi\bar{\chi}\gamma^{5}\chi-h_1\phi\bar e e-ih_2\phi\bar{e}\gamma^5e\,, 
\label{eq:Lfs}
\end{eqnarray}
where $\lambda_1$ and $\lambda_2$ are coupling constants, while $\chi$ is the Dirac spinor that describes the DM particle.~Alternatively, in the case of a vector mediator, $G_\mu$, we assume the following Lagrangian
\begin{eqnarray}
\mathcal{L}_{\chi Ge} &=& \ii\bar{\chi}\slashed{D}\chi - m_\chi\bar{\chi}\chi \nonumber\\
&-&\frac{1}{4}\mathcal{G}_{\mu\nu}\mathcal{G}^{\mu\nu}+\frac{1}{2}m_{G}^2G_{\mu}G^{\mu}-\frac{\lambda_G}{4}(G_\mu G^\mu)^2\nonumber\\
&+& \ii\bar{e}\slashed{D} e - m_e \bar e e \nonumber\\
&-&\lambda_{3}\bar\chi\gamma^\mu\chi G_{\mu}-\lambda_{4}\bar\chi\gamma^\mu\gamma^5\chi G_{\mu}\nonumber\\
&-&h_3\bar{e}\gamma_{\mu}eG^{\mu}-h_4\bar{e}\gamma_{\mu}\gamma^{5}eG^{\mu}\,,
\label{eq:Lfv}
\end{eqnarray}
where $h_3$ and $h_4$ are additional coupling constants, and the remaining symbols were introduced above.

\subsection{Vector dark matter}
\label{sec:vector}
Finally, we focus on the case of spin-1 DM particles with mass $m_X$ described by a complex vector field, $X_\mu$.~For DM-electron interactions mediated by a scalar field, $\phi$, we assume the Lagrangian,
\begin{eqnarray}
\mathcal{L}_{X\phi e}&=&-\frac{1}{2}{\mathcal{X}}_{\mu\nu}^{\dagger}\mathcal{X}^{\mu\nu}+m_{X}^2X_{\mu}^{\dagger}X^{\mu}-\frac{\lambda_{X}}{2}(X_{\mu}^{\dagger}X^{\mu})^2 \nonumber \\
&+&\frac{1}{2}(\partial_{\mu}\phi)^2-\frac{1}{2}m_{\phi}^2\phi^2-\frac{m_\phi \mu_1}{3}\phi^3-\frac{\mu_2}{4}\phi^4 \nonumber\\
&+&\ii\bar{e}\slashed{D}e-m_{e}\bar{e}e \nonumber\\
&-&b_1m_X\phi X_{\mu}^{\dagger}X^{\mu}-\frac{b_{2}}{2}\phi^2X_{\mu}^{\dagger}X^{\mu}  \nonumber\\ 
&-&h_1\phi\bar{e}e-ih_2\phi\bar{e}\gamma^{5}e\,,  
\end{eqnarray}
where $\lambda_X$, $b_1$ and $b_2$ are new coupling constants, and $\mathcal{X}_{\mu\nu}$ is the field strength tensor for $X_\mu$.~On the other hand, in the case of DM-electron interactions mediated by a spin-1 particle, we assume
\begin{eqnarray}
\mathcal{L}_{XGe}&=& -\frac{1}{2}\mathcal{X}^{\dagger}_{\mu\nu}\mathcal{X}^{\mu\nu}+m_{X}^2X^{\dagger}_{\mu}X^{\mu}-\frac{\lambda_{X}}{2}(X_{\mu}^{\dagger}X^{\mu})^2 \nonumber\\
&-&\frac{1}{4}\mathcal{G}_{\mu\nu}\mathcal{G}^{\mu\nu}+\frac{1}{2}m_{G}^2G_\mu G^\mu-\frac{\lambda_G}{4}(G_\mu G^\mu)^2\nonumber \\
&+&i\bar{e}\slashed{D}e-m_{e}\bar{e}e\nonumber\\
&-&\frac{b_3}{2}G_\mu G^\mu (X^{\dagger}_\nu X^{\nu}) -\frac{b_{4}}{2}(G^{\mu}G^{\nu})(X^{\dagger}_{\mu}X_{\nu}) \nonumber\\ 
&-&\left[ ib_{5}X_{\nu}^{\dagger}\partial_{\mu}X^{\nu}G^\mu + b_{6}X_{\mu}^{\dagger}\partial^\mu X_{\nu}G^{\nu}\right.\nonumber\\ 
&+& \left.b_{7}\epsilon_{\mu\nu\rho\sigma}(X^{\dagger\mu}\partial^{\nu}X^{\rho})G^{\sigma} +h.c.\right]\nonumber\\
&-&h_3G_\mu\bar{e}\gamma^\mu e - h_4 G_\mu\bar{e}\gamma^\mu\gamma^{5}e\,,
\end{eqnarray}
where $b_3$, $b_4$, $b_5$, $b_6$ and $b_7$ are coupling constants.~Contrary to all other coupling constants, which can without loss of generality be assumed to be real, $b_6$ and $b_7$ are in general complex variables~\cite{Catena:2017xqq}.

\section{Expected rate of electron transitions in detector materials}
\label{sec:dd}
From the Lagrangians introduced in Sec.~\ref{sec:lagrangians}, we now calculate the expected rate of electron transitions induced by DM scattering in isolated atoms and crystals.~We start by reviewing the results of~\cite{Catena:2019gfa,Catena:2021qsr} (Sec.~\ref{sec:gen}).~We then apply these results to the general models of Sec.~\ref{sec:lagrangians} (Sec.~\ref{sec:app}).

\subsection{General considerations}
\label{sec:gen}
Let us consider the scattering of a DM~particle of mass~$m_{\rm DM}$, initial velocity in the detector rest frame $\vec v$, and momentum~$\vec p = m_{\rm DM} \vec v$ by an electron bound to an isolated atom or to a crystal, and described by the initial state~$\ket{\mathbf{e}_1}$.~Furthermore, let us denote by $\ket{\mathbf{e}_2}$ the state of the outgoing electron, and by $\vec p^\prime$ the momentum of the final state DM particle.~The momentum transfer in the scattering is thus $\vec q = \vec p-\vec p^\prime$, and the rate for the  $\ket{\mathbf{e}_1}$ to $\ket{\mathbf{e}_2}$  transition is given by~\cite{Catena:2019gfa,Catena:2021qsr}
\begin{align}
\mathscr{R}_{1\rightarrow 2}&=\frac{n_{\rm DM}}{16 m^2_{\rm DM} m^2_e} \,
\int \frac{{\rm d}^3 q}{(2 \pi)^3} \int {\rm d}^3 \vec v\, f(\vec v)  
(2\pi) \delta(E_f-E_i) \overline{\left| \mathcal{M}_{1\rightarrow 2}\right|^2}\, , 
\label{eq:transition rate}
\end{align}
where $E_i$ and $E_f$ are the total initial and final energy, respectively.~They can be written as follows
\begin{align}
    E_i &= m_{\rm DM} + m_e + \frac{|\vec p|^2}{2m_{\rm DM}}+E_1 \nonumber\\
    E_f &= m_{\rm DM} + m_e + \frac{|\vec p-\vec q|^2}{2m_{\rm DM}}+E_2\, .
\end{align}
Here, $E_1<0$ ($E_2>0$) is the energy eigenvalue of the state~$\ket{\mathbf{e}_1}$ ($\ket{\mathbf{e}_2}$).~This convention for the signs of $E_1$ and $E_2$ reflects our interest in transitions from bound atomic (valence) to free (conduction) states (see Secs.~\ref{sec:atoms} and~\ref{sec:crystals}).~In Eq.~(\ref{eq:transition rate}), $\mathcal{M}_{1\rightarrow 2}$ is the electron transition amplitude, defined as the overlap integral,
\begin{align}
     \mathcal{M}_{1\rightarrow 2} (\vec q, \vec v) &=
     \int  
     \frac{{\rm d}^3 \boldell}{(2 \pi)^3} \, \widetilde{\psi}_2^*(\boldsymbol{\ell}+\vec q)  \,
\mathcal{M}(\vec q,\vec v^\perp_{\rm el}(\vec v,\vec q,\boldsymbol{\ell})) \,
\widetilde{\psi}_1(\boldsymbol{\ell}) \, ,
\label{eq:transition amplitude}
\end{align}
where $\widetilde{\psi}_1$ ($\widetilde{\psi}_2$) is the Fourier transform of $\psi_1$ ($\psi_2$), while $\psi_1$ ($\psi_2$) is the wave function associated with the state $\ket{\mathbf{e}_1}$ ($\ket{\mathbf{e}_2}$),
and $\mathcal{M}$ is the free electron scattering amplitude.~In the {\it non-relativistic}, elastic scattering of a DM particle by free electrons, $\mathcal{M}$ uniquely depends on the momentum transfer $\vec q$ and on the transverse relative velocity $\vec v^\perp_{\rm el}=\vec v- \vec q/(2\mu_{\rm DM e}) -\boldsymbol{\ell}/m_e$, where $\mu_{\rm DM e}$ is the DM-electron reduced mass and $\boldsymbol{\ell}$ is the initial state electron momentum.~This simplification is a consequence of Galilean invariance (that is, the invariance under constant shifts of particle velocities) and three-dimensional momentum conservation, which allow one to express $\mathcal{M}$ as a function of two kinematic variables only (see~\cite{Dobrescu:2006au,Fan:2010gt,Fitzpatrick:2012ix,Catena:2019gfa,Catena:2021qsr} for further details).~Notice that Galilean invariance is fulfilled when both the DM particle and the electron are non-relativistic.~We can now expand $\mathcal{M}$ at linear order in $\boldsymbol{\ell}/m_e$, which implies~\cite{Catena:2019gfa,Catena:2021qsr},
\begin{align}
    \mathcal{M}_{1\rightarrow 2}(\vec q, \vec v)  &\simeq \left. \mathcal{M}(\vec q,\vec v^\perp_{\rm el}) \right|_{\boldell=\mathbf{0}} f_{1\rightarrow 2}(\vec q) 
    + m_e \left.\nabla_{\boldell} \mathcal{M}(\vec q,\vec v^\perp_{\rm el})  \right|_{\boldell=\mathbf{0}} \cdot \vec f_{1\rightarrow 2}(\vec q) \,,
    \label{eq:M_expansion}
\end{align}
where in the right-hand-side we introduced the scalar and vectorial wave function overlap integrals,
\begin{align}
       f_{1\rightarrow 2}(\vec q) &\equiv \int\dd^3 \textbf{x}\, \psi_2^*(\mathbf{x})\,e^{i\vec q\cdot\mathbf{x}}\,\psi_1\left( \mathbf{x} \right)\, , \label{eq: f scalar}\\
    \vec f_{1\rightarrow 2}(\vec q) &\equiv \int\dd^3 \textbf{x}\, \psi_2^*(\mathbf{x})\,e^{i\vec q\cdot\mathbf{x}}\,\frac{i\nabla_{\mathbf{x}}}{m_e}\psi_1\left( \mathbf{x} \right) \, . \label{eq: f vectorial}
\end{align}
In Eq.~(\ref{eq:transition rate})$, n_{\rm DM} = \rho_{\rm DM} / m_{\rm DM}$ is the local DM~number density, $\rho_{\rm DM} = 0.4\,\text{GeV cm}^{-3}$ the local DM mass density~\cite{Catena:2009mf}, and~$f(\vec v)$ the local DM velocity distribution boosted to the detector rest frame.~For $f(\vec v)$, we assume a truncated Maxwell-Boltzmann distribution with local standard of rest speed $v_0=~238$~km~s$^{-1}$~\cite{Baxter:2021pqo}, galactic escape speed $v_\mathrm{esc}=544$~km~s$^{-1}$~\cite{Baxter:2021pqo} and Earth's speed in the galactic frame (where the mean DM particle velocity is zero) $v_e=250.5$ km~s$^{-1}$~\cite{Baxter:2021pqo}.

The {\it theoretical} rate of DM-induced electron transitions in a detector material, $\mathscr{R}_{\rm theory}$, is related to the rate for state $\ket{\mathbf{e}_1}$  to state $\ket{\mathbf{e}_2}$ transitions, $\mathscr{R}_{1 \rightarrow 2}$, by a sum over all ``relevant'' initial and final states.~For detectors that can be approximated by a system of isolated atoms, and that search for atomic ionisations induced by the scattering of DM particles in the detector, the relevant initial states are bound states characterised by the principal, angular and magnetic quantum numbers, $n$, $\ell$ and $m$, respectively.~The relevant final states are free electron states characterised by an ``asymptotic'' momentum $k'$, and by the electron angular and magnetic quantum numbers, $\ell'$ and $m'$, respectively.~We refer to the appendices of~\cite{Catena:2019gfa} for an explicit form of the corresponding wave functions.~By replacing $1$ with the quantum numbers $\{n\ell m\}$, and $2$ with $\{k^{\prime}\ell^{\prime}m^{\prime}\}$ in Eq.~(\ref{eq:M_expansion}), and inserting Eq.~(\ref{eq:M_expansion}) into Eq.~(\ref{eq:transition rate}), we obtain
\begin{align}
\mathscr{R}_{\rm theory} = \mathscr{R}_{\rm ion} &\equiv 
2 \sum_{n,\ell}\sum_{m=-\ell}^\ell \sum_{\ell^\prime=0}^\infty\sum_{m^\prime=-\ell^\prime}^{\ell^\prime}\int\dd k' \frac{V k^{\prime 2}}{(2\pi)^3} \mathscr{R}_{n\ell m\rightarrow k^{\prime}\ell^{\prime}m^{\prime}}\, ,
\label{eq:rateobs1}
\end{align}
where the $(\ell,m)$ sum runs over the outermost occupied orbitals, e.g. the 4s, 4p, 4d, 5s, and 5p orbitals in the case of xenon atoms, which is of interest for the present work.~In Eq.~(\ref{eq:rateobs1}) we assume that wave functions have been normalised over a finite volume $V$, and the overall factor of 2 accounts for the double occupation of each electronic state due to spin degeneracy. For crystal detectors that search for electronic transitions from valence to conduction bands in semiconductor targets, the relevant initial states are occupied valence states labeled by a band index $i$ and a reciprocal lattice momentum in the first Brillouin Zone (BZ), $\vec k$.~Similarly, the relevant final states are in this case unoccupied conduction states characterised by a band index $i'$ and a reciprocal lattice momentum in the first BZ, $\vec k'$.~For crystal detectors, we thus find 
\begin{align}
\mathscr{R}_{\rm theory} = \mathscr{R}_{\rm crystal} \equiv 2 \sum_{i i'} \int_{\rm BZ} \frac{V {\rm d}^3 k}{(2\pi)^3} \int_{\rm BZ} \frac{V {\rm d}^3 k'}{(2\pi)^3} \,
\mathscr{R}_{i\vec k \rightarrow i'\vec k'} \,.
\end{align}
We refer to~\cite{Catena:2021qsr} for an explicit form of the Bloch wave functions $\psi_{i\vec k}$ and $\psi_{i'\vec k'}$ entering Eqs.~(\ref{eq: f scalar}) and (\ref{eq: f vectorial}).

In general, we express the rate $\mathscr{R}_{\rm theory}$ within our ``response function'' formalism~\cite{Catena:2019gfa,Catena:2021qsr}, where 
\begin{align}
\mathscr{R}_{\rm theory}=\frac{n_{\rm DM} 
}{128\pi m_{\rm DM}^2 m_e^2}&\int \mathrm{d} (\ln\Delta E)\int \mathrm{d} q \, q \,\widehat{\eta}\left(q, \Delta E
\right)
\sum_{l=1}^r \Re\left[\mathcal{R}_l^*(q,v) \overline{\mathcal{W}}_l(q,\Delta E)\right]\,.
\label{eq:Robs}
\end{align}
With an appropriate choice of ``material response'' functions $\overline{\mathcal{W}}_l(q,\Delta E)$, and ``DM response'' functions $\mathcal{R}_l(q,v)$, Eq.~(\ref{eq:Robs}) applies to any detector material and particle physics model.~In particular, it applies to detectors that can be approximated by a set of isolated atoms, such as XENON10 and XENON1T, and to crystal detectors, such as SENSEI and EDELWEISS.~The velocity operator in Eq.~(\ref{eq:Robs}), $\widehat{\eta}\left(q, \Delta E \right)$, depends on the momentum transfer, $\vec q$, and on the deposited energy, $\Delta E$.~It linearly acts on the generic function of velocities, $g(\vec v)$, as shown below
\begin{align}
\widehat{\eta}(q, \Delta E) \left[ g(\vec v) \right] &=  \int_{|\vec v|\ge v_{\rm min}
}
{\rm d}^3v \, g(\vec v)
\frac{f_\chi(\vec v)}{v} \,,
\end{align}
where 
\begin{align}
v_{\rm min}=\frac{q}{2 m_{\rm DM}} + \frac{\Delta E}{q } \,,
\label{eq:vmin}
\end{align}
is the minimum velocity required to transfer a momentum $\vec q$ when the deposited energy is $\Delta E$, and $q\equiv |\vec q|$.~Below, we first provide explicit expressions for the $\overline{\mathcal{W}}_l(q,\Delta E)$, $l=1,\dots r$ functions that characterise detectors consisting of isolated atoms (Sec.~\ref{sec:atoms}) and crystals (Sec.~\ref{sec:crystals}).~We then calculate the functions $\mathcal{R}_l(q,v)$, $l=1,\dots r$, for all models in Sec.~\ref{sec:lagrangians} (Sec.~\ref{sec:app}).

\subsubsection{Isolated atoms}
\label{sec:atoms}
The material response functions $\overline{\mathcal{W}}_l$ are defined via an integral over the azimuthal and polar angles of $\vec q$,
\begin{align}
    \overline{\mathcal{W}}_l(q,\Delta E)\equiv&\int \mathrm{d}\Omega_q\, \mathcal{W}_l(\vec q,\Delta E) \,,
    \label{eq:omega}
\end{align}
where, for detectors that can be modelled as a sample of isolated atoms, such as XENON10 or XENON1T, 
\begin{align}
    \mathcal{W}_l(\vec q,\Delta E)&=\frac{2}{\pi} \Delta E  \sum_{n,\ell,m} \sum_{\ell',m'}  
   \int \frac{V k^{\prime 2} \,\mathrm{d}k^\prime}{(2\pi)^3} \,\mathscr{B}_l \, 
   \delta(\Delta E - E_{k'\ell' m'} + E_{n\ell m} )
    \label{eq:Watoms}\,,
\end{align}
and
\begin{align}
\mathscr{B}_1 =& \left| f_{n\ell m \rightarrow k'\ell' m'} \right|^2 \\
\mathscr{B}_2=&\frac{\vec q}{m_e} \cdot(f_{n\ell m \rightarrow k'\ell' m'}) (\vec f_{n\ell m \rightarrow k'\ell' m'})^* \\
\mathscr{B}_3=&\left| \vec f_{n\ell m \rightarrow k'\ell' m'}\right|^2  \\
\mathscr{B}_4=& \left|\frac{\vec q}{m_e} \cdot \vec f_{n\ell m \rightarrow k'\ell' m'} \right|^2 \,.
\end{align}
$E_{n\ell m}$ ($E_{k'\ell' m'}$) is the energy of the initial (final) state electron and $f_{n\ell m \rightarrow k'\ell' m'}$ ($\vec f_{n\ell m \rightarrow k'\ell' m'}$) replaces $f_{1\rightarrow 2}$ ($\vec f_{1\rightarrow 2}$).~To compare the results reported here with those in~\cite{Catena:2019gfa}, we note that the $W_l^{n\ell}(k',q)$ functions introduced in~\cite{Catena:2019gfa} and the $\overline{\mathcal{W}}_l(q,\Delta E)$ functions defined here are related by
\begin{align}
\overline{\mathcal{W}}_l(q,\Delta E) = 2 \Delta E \sum_{n,\ell} \int \frac{{\rm d}k'}{k'} \, W_l^{n\ell}(k',q) \, \delta(\Delta E - E_{k'\ell' m'} + E_{n\ell m})\,.
\label{eq:newold}
\end{align}
 Since we are interested in the ionisation of isolated atoms, in Eq.~(\ref{eq:newold})  $E_{k'\ell' m'}=k^{\prime 2}/(2 m_e)$ and 
  the initial state energy, $E_{n\ell m}$, does not depend on $m$.~Consequently, Eq.~(\ref{eq:newold}) consistently gives a response function $\overline{\mathcal{W}}_l(q,\Delta E)$ that is independent of quantum numbers.~Also, requiring that $v_{\rm min}(q,\Delta E)$ is less than the maximum speed a DM particle can have in the galaxy, $v_{\rm max}=v_e+v_{\rm esc}$,  implies that only certain regions in the $(q,\Delta E)$ plane are kinematically allowed.
 
 We conclude this subsection by providing an explicit relation between our response function $\overline{\mathcal{W}}_1$ and the standard ionisation form factor $f^{n\ell}_{\rm ion}$ used within the dark photon model, namely
\begin{align}
\overline{\mathcal{W}}_1(q,\Delta E) = 2 \Delta E \sum_{n,\ell} \int \frac{{\rm d}k'} {k'}\, |f_{\rm ion}^{n\ell}(k',q)|^2 \, \delta(\Delta E - E_{k'\ell' m'} + E_{n\ell m})\,.
\label{eq:newold2}
\end{align}
In the numerical applications, we compute the response functions $W_l^{n\ell}(k',q)$ by using the {\sffamily DarkART} code~\cite{DarkART}, which implements the equations we derived in~\cite{Catena:2019gfa}.

\subsubsection{Crystals}
\label{sec:crystals}
Similarly, in the case of crystal detectors such as SENSEI and EDELWEISS one can use Eq.~(\ref{eq:omega}) with 
\begin{align}
    \mathcal{W}_l(\vec q,\Delta E)&=\frac{2}{\pi} \Delta E \sum_{ii^\prime}\int_{\textrm{BZ}}\frac{V \mathrm{d}^3k }{(2\pi)^3}\int_{\textrm{BZ}}\frac{V \mathrm{d}^3k^\prime}{(2\pi)^3} \,\mathscr{B}_l \, 
   \delta(\Delta E -E_{i'\vec k^\prime} + E_{i\vec k})
    \label{eq:Wcrystals}\,
\end{align}
to calculate the response functions $\overline{\mathcal{W}}_l(q,\Delta E)$.~In this case, $r=5$ functions $\mathscr{B}_l$ are non zero~\cite{Catena:2021qsr},
\begin{align}
\mathscr{B}_1 =& \left| f_{i\vec k\rightarrow i^\prime \vec k^\prime} \right|^2 \\
\mathscr{B}_2=&\frac{\vec q}{m_e} \cdot(f_{i\vec k\rightarrow i^\prime\vec k^\prime}^\prime) (\vec f_{i\vec k\rightarrow i^\prime \vec k^\prime})^* \\
\mathscr{B}_3=&\left| \vec f_{i\vec k\rightarrow i^\prime \vec k^\prime} \right|^2  \\
\mathscr{B}_4=& \left|\frac{\vec q}{m_e} \cdot \vec f_{i\vec k\rightarrow i^\prime \vec k^\prime} \right|^2 \\
\mathscr{B}_5=& i\frac{\vec q}{m_e} \cdot \left[\vec f_{i\vec k\rightarrow i^\prime \vec k^\prime} \times \left(\vec f_{i\vec k\rightarrow i^\prime\vec k^\prime}\right)^*\right] \,.
\end{align}
Interestingly, the response function induced by $\mathscr{B}_5$ is identically zero in the case of isolated atoms~\cite{Catena:2019gfa}.~Here, $E_{i\vec k}$ ($E_{i'\vec k'}$) is the energy of the $i\vec k$ ($i'\vec k'$)  initial state valence (final state conduction) electron, while $f_{1\rightarrow 2}$ ($\vec f_{1\rightarrow 2}$) is replaced by $f_{i\vec k}$ ($\vec f_{i'\vec k'}$), and $V=N_{\rm cell}V_{\rm cell}$, where $V_{\rm cell}$ is the volume of a unit cell, and $N_{\rm cell}$ the number of unit cells in the crystal.~Note that the functions $\overline{\mathcal{W}}_l$ defined via Eq.~(\ref{eq:Wcrystals}) and the functions $\overline{W}_l$ introduced in~\cite{Catena:2021qsr} are related as follows
\begin{align}
\overline{\mathcal{W}}_l(q,\Delta E)=N_{\rm cell} \, \overline{W}_l(q,\Delta E)\,.
\end{align}
By absorbing $N_{\rm cell}$ in the definition of $\overline{\mathcal{W}}_l$, Eq.~(\ref{eq:Robs}) applies to both isolated atoms and crystals.~Finally, $\overline{\mathcal{W}}_1$ and the crystal form factor used in the dark photon model, $f_{\text {crystal}}$, are related by
\begin{equation}
  \overline{\mathcal{W}}_1(q,\Delta E) = \frac{8\Delta E\alpha m_e^2 N_{\rm cell}}{q^3}  \left|f_{\text {crystal}}\left(q, \Delta E\right)\right|^{2}\, , \label{eq:Comparison_Essig}
\end{equation}
where $\alpha$ is the fine structure constant.~In the numerical applications, we compute the response functions $\overline{W}_l(q,\Delta E)$ by using the {\sffamily QEdark-EFT} code~\cite{Urdshals2021May}, which implements the equations we derived in~\cite{Catena:2021qsr}.

\subsubsection{Unified notation}
\begin{figure}[t]
\begin{center}
\begin{minipage}[t]{0.495\linewidth}
\centering
\includegraphics[width=\textwidth]{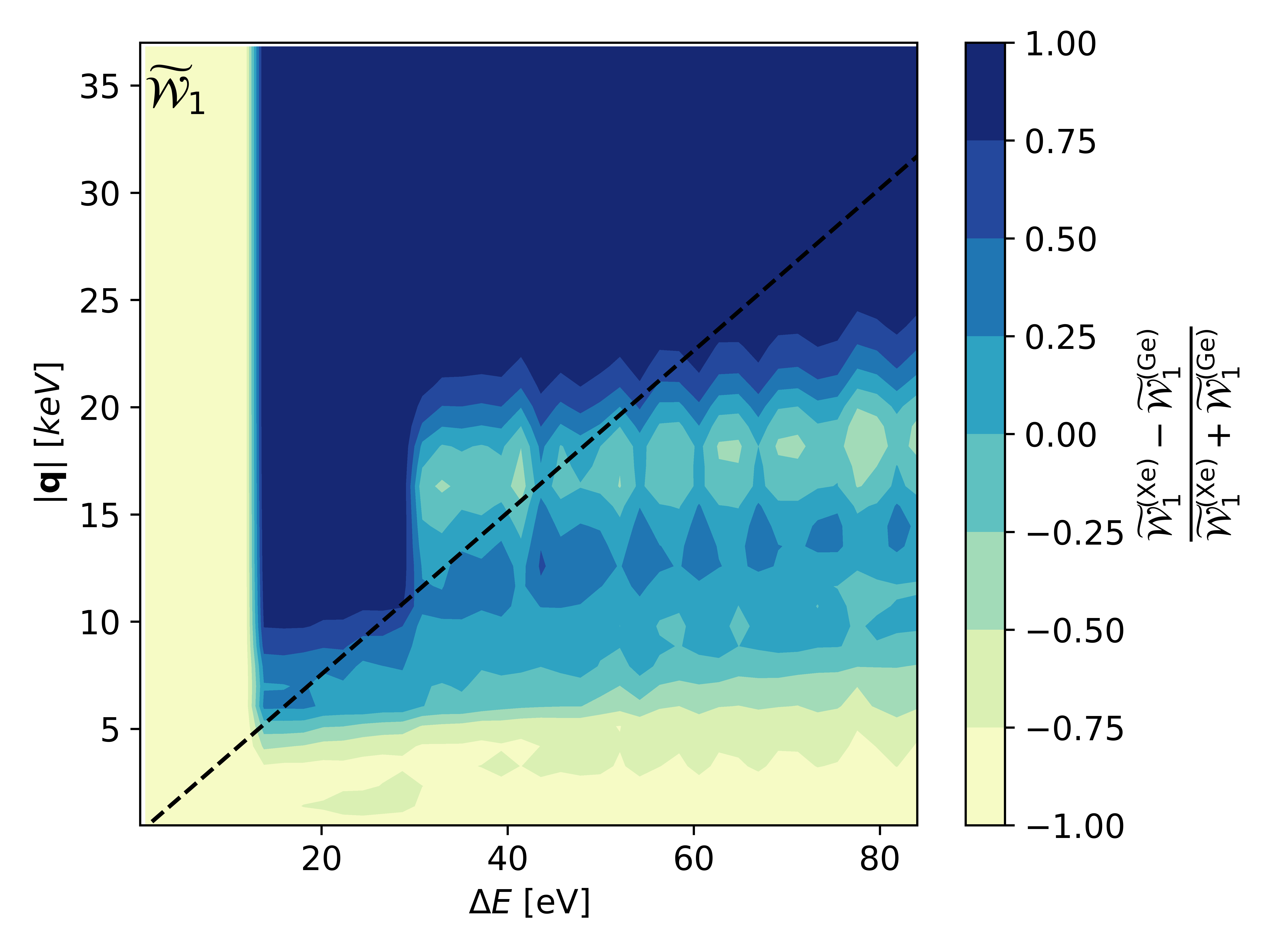}
\end{minipage}
\begin{minipage}[t]{0.495\linewidth}
\centering
\includegraphics[width=\textwidth]{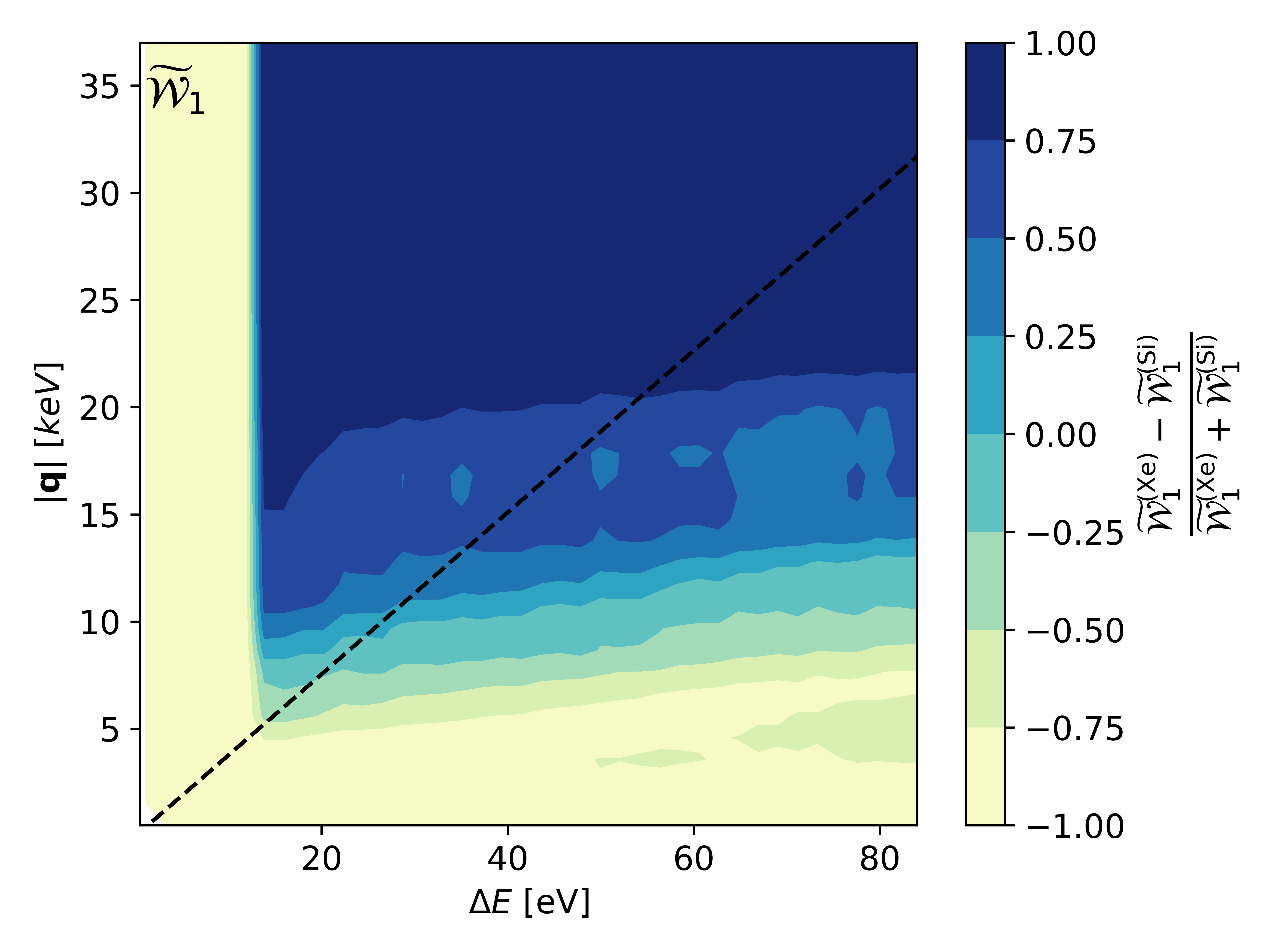}
\end{minipage}
\begin{minipage}[t]{0.495\linewidth}
\centering
\includegraphics[width=\textwidth]{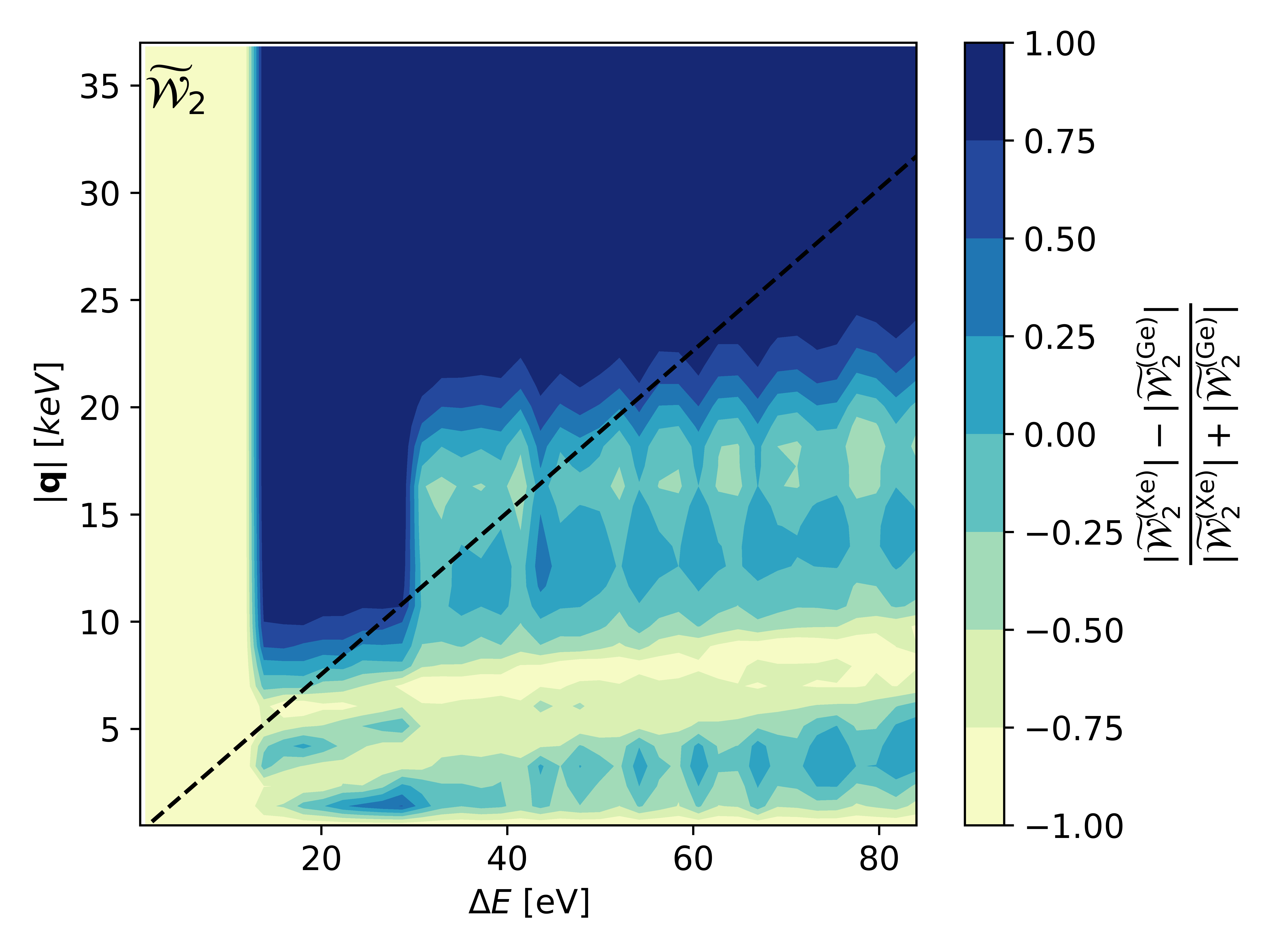}
\end{minipage}
\begin{minipage}[t]{0.495\linewidth}
\centering
\includegraphics[width=\textwidth]{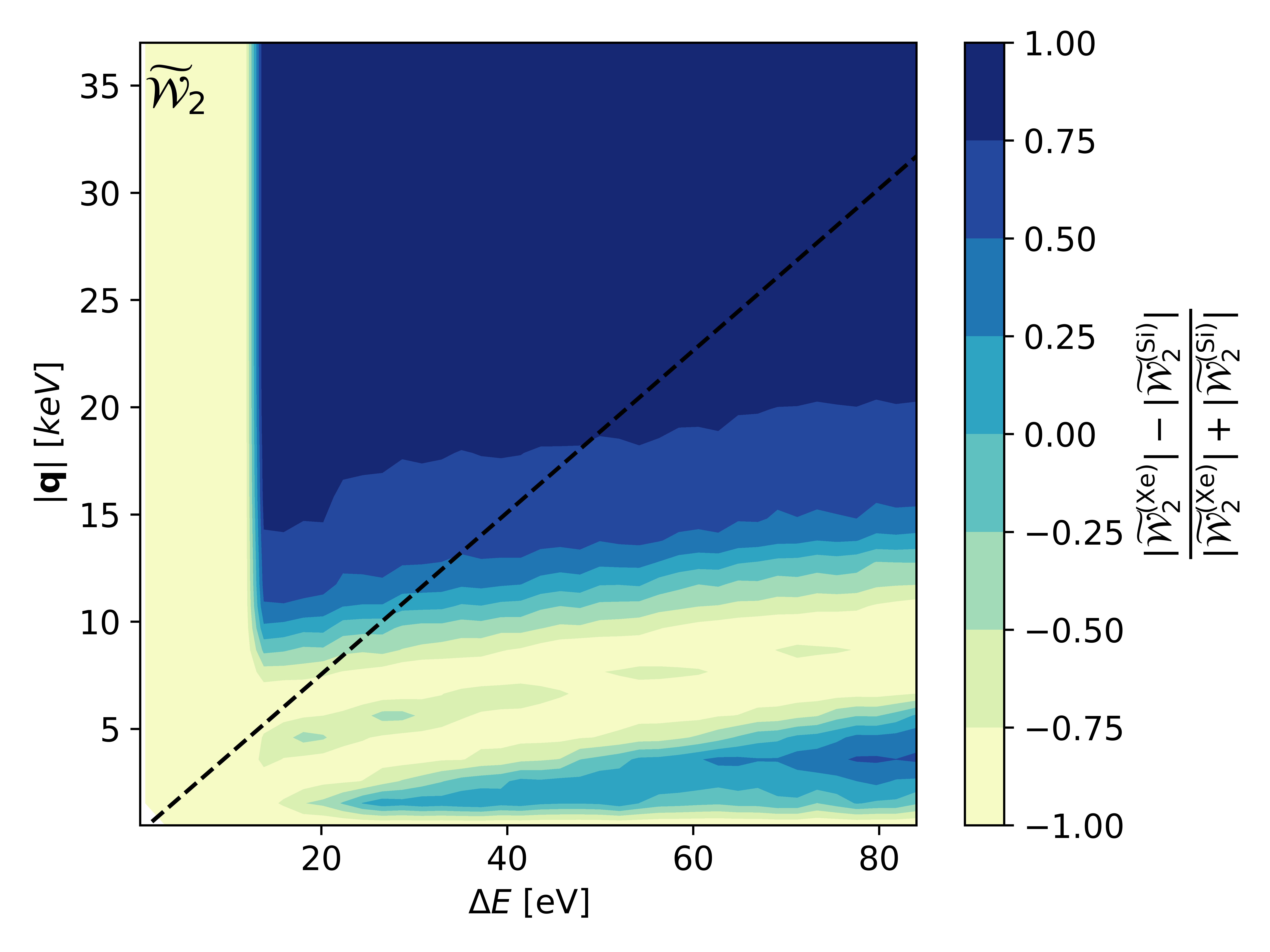}
\end{minipage}
\end{center}
\caption{Relative differences between detector response functions per unit mass for xenon and germanium (left), and for xenon and silicon (right) detectors in the $(q,\Delta E)$ plane.~The top (bottom) panels refer to $l=1$ ($l=2$).~Regions above the black dashed line are kinematically allowed.}
\label{fig:Wratio1}
\end{figure}

\begin{figure}[t]
\begin{center}
\begin{minipage}[t]{0.495\linewidth}
\centering
\includegraphics[width=\textwidth]{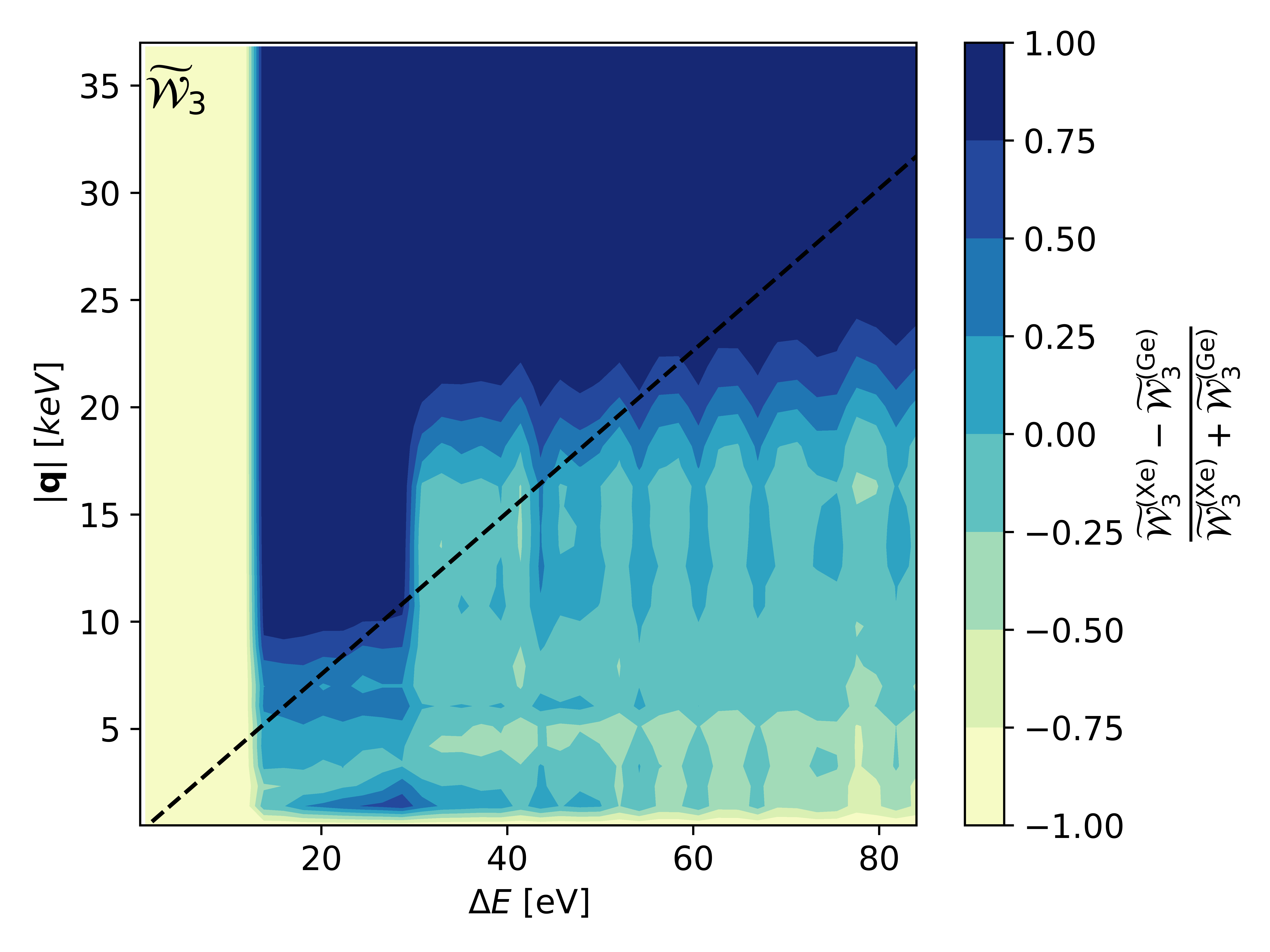}
\end{minipage}
\begin{minipage}[t]{0.495\linewidth}
\centering
\includegraphics[width=\textwidth]{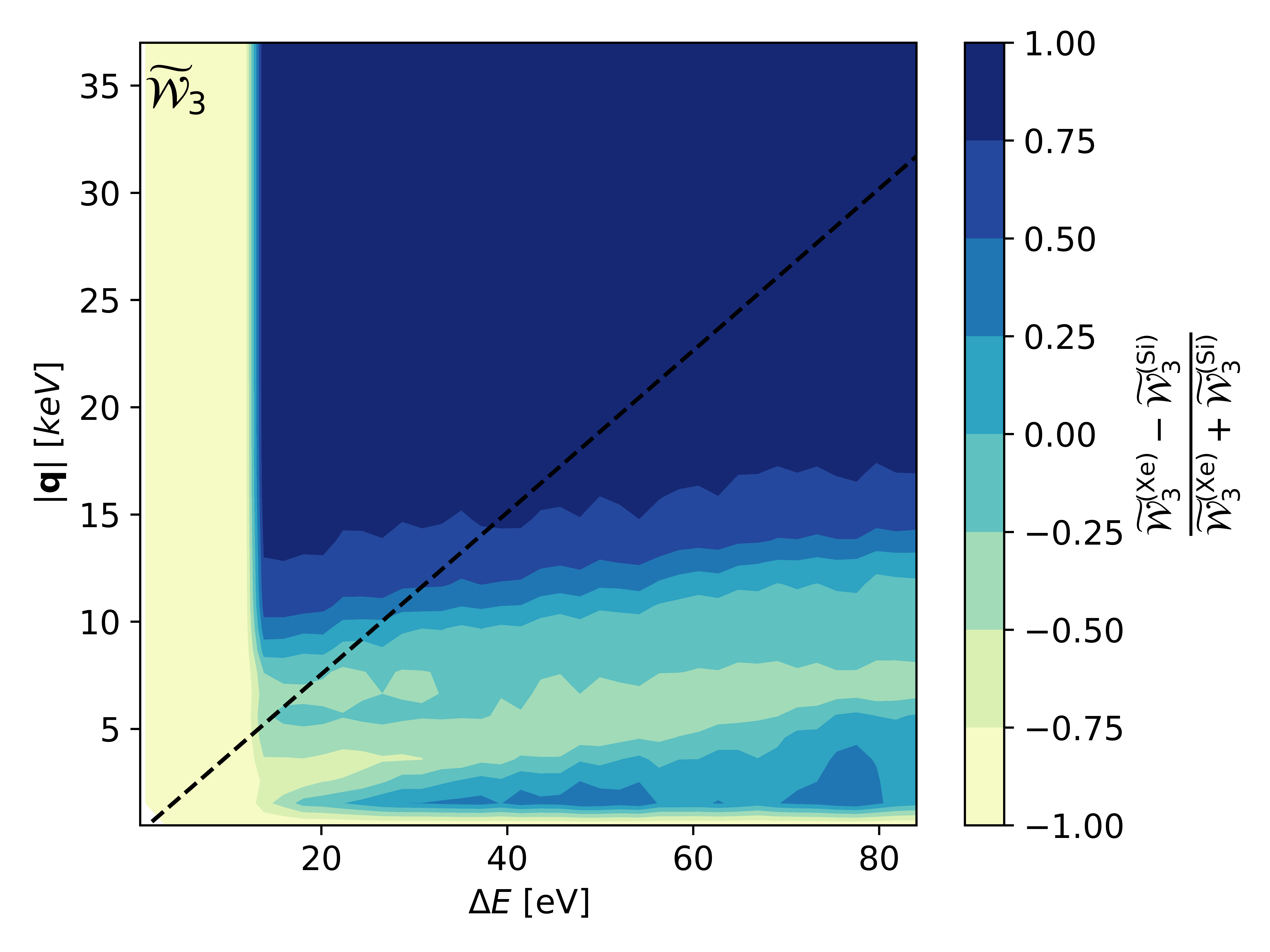}
\end{minipage}
\begin{minipage}[t]{0.495\linewidth}
\centering
\includegraphics[width=\textwidth]{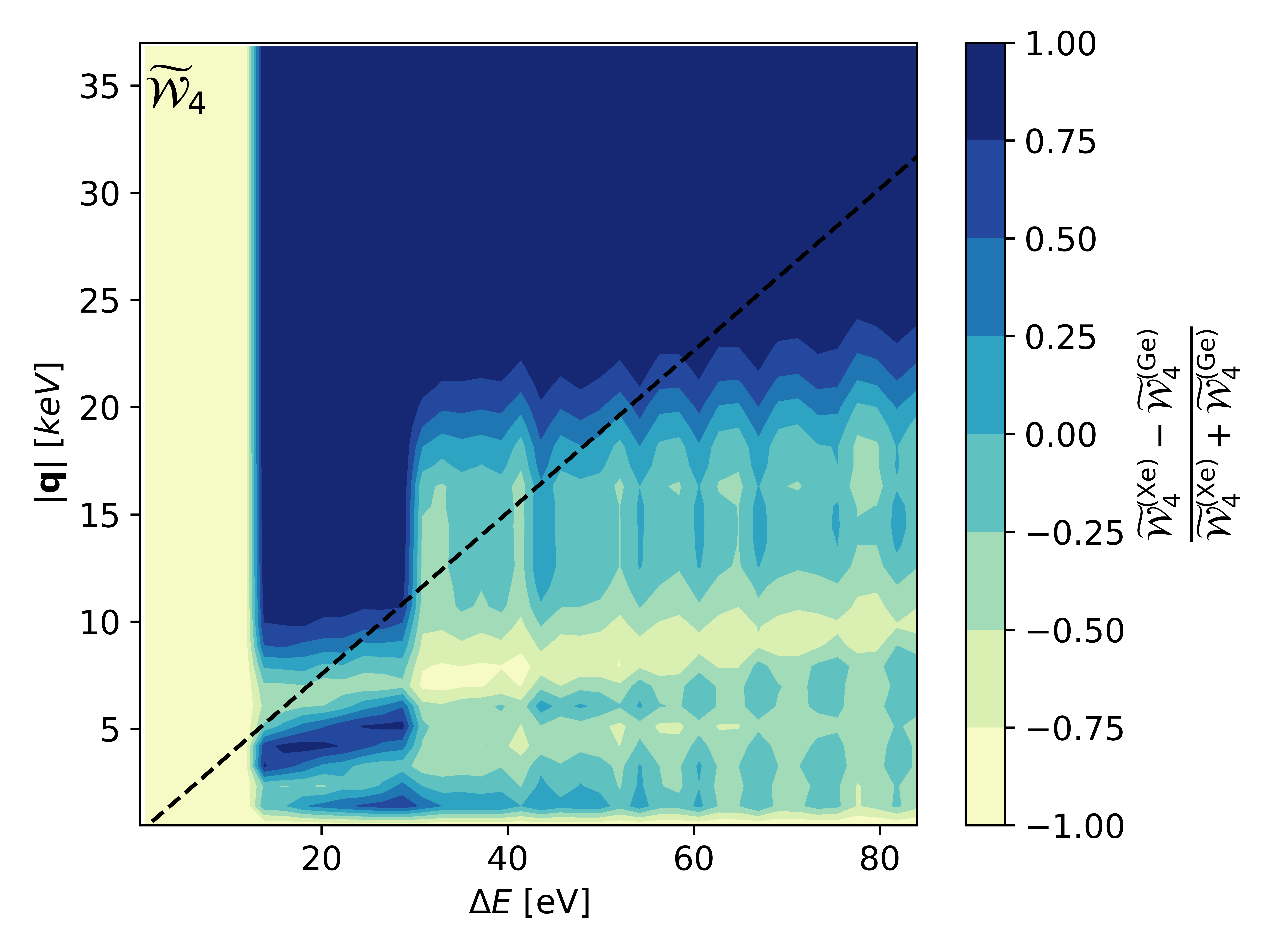}
\end{minipage}
\begin{minipage}[t]{0.495\linewidth}
\centering
\includegraphics[width=\textwidth]{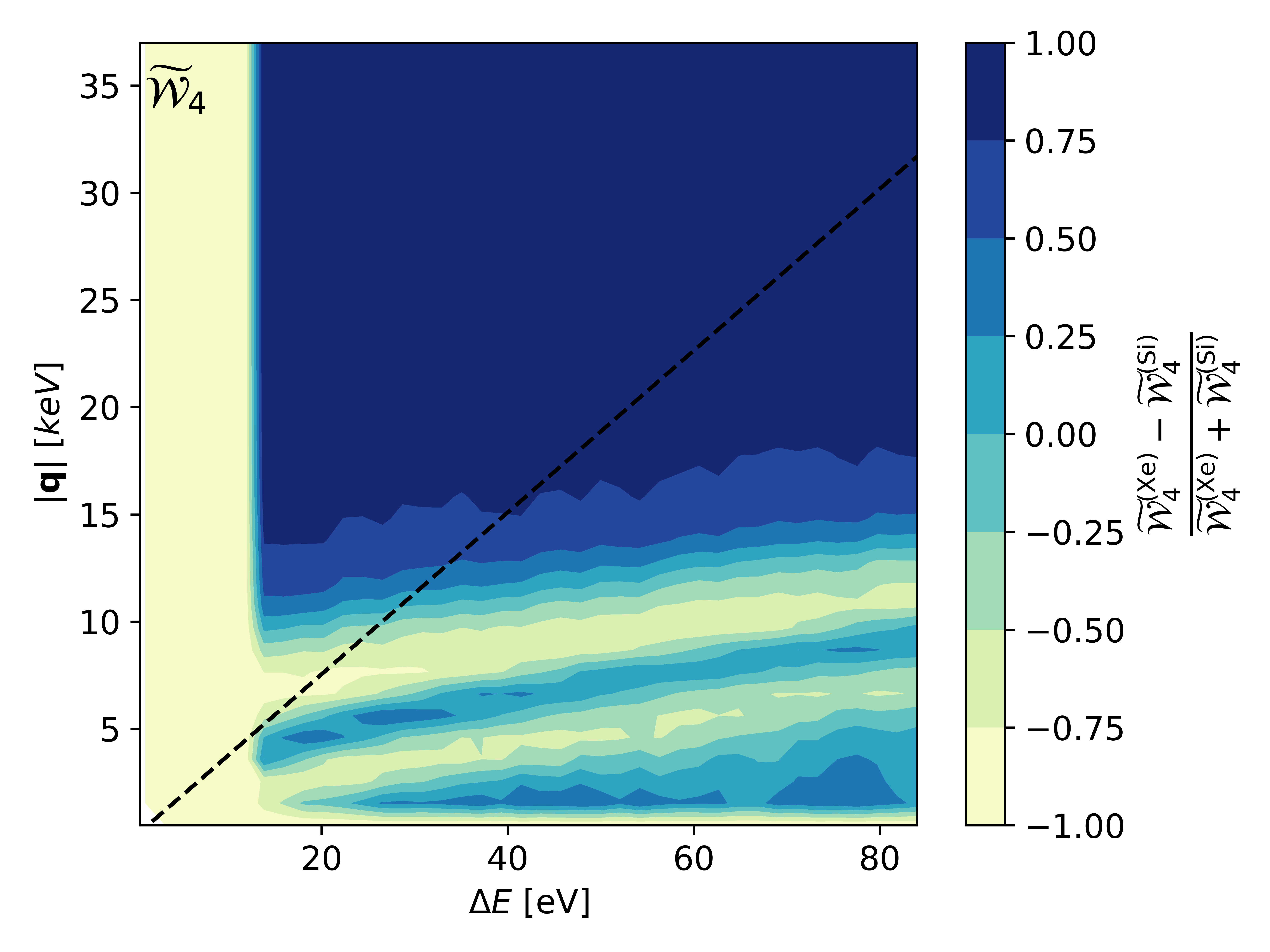}
\end{minipage}
\end{center}
\caption{The same as for Fig.~\ref{fig:Wratio1}, now for $l=3$ and $l=4$.}
\label{fig:Wratio2}
\end{figure}

Remarkably, the response functions $\mathcal{W}_l(\vec q,\Delta E)$ for isolated atoms, Eq.~(\ref{eq:Watoms}), and those for crystals, Eq.~(\ref{eq:Wcrystals}), have the very same form.~Specifically, they can both be expressed  as follows
\begin{align}
    \mathcal{W}_l(\vec q,\Delta E)&=\frac{2}{\pi} \Delta E \sum_{\{1\},\{2\}} \mathscr{B}_l \, 
   \delta(\Delta E -E_{2}+E_{1})
    \label{eq:Wcompact}\,,
\end{align}
where $ \sum_{\{1\},\{2\}}$ represents the sums/integrals over all relevant physical states in Eq.~(\ref{eq:Watoms}) and Eq.~(\ref{eq:Wcrystals}).~In the case of isolated atoms, Eq.~(\ref{eq:Wcompact}) gives the ``detector response'' for a single atom, whereas in the case of crystal detectors Eq.~(\ref{eq:Wcompact}) gives the ``detector response'' for a single crystal.~In order to compare the two quantities, we introduce a {\it detector response per unit mass},
\begin{align}
\widetilde{\mathcal{W}}_l(q,\Delta E) \equiv \bar{\mathcal{W}}_l(q,\Delta E)/\widetilde{m}\,,
\end{align} 
where for isolated atoms $\widetilde{m}=m_{\rm Xe}$, $m_{\rm Xe}$ being the mass of a xenon atom, whereas for crystals, $\widetilde{m}=m_{\rm cell} N_{\rm cell}$, $m_{\rm cell}$ and $N_{\rm cell}$ being the mass and number of unit cells. Fig.~\ref{fig:Wratio1} shows the ratios
\begin{align}
(\widetilde{\mathcal{W}}_l^{(\rm Xe)}-\widetilde{\mathcal{W}}_l^{(\rm Ge)})/(\widetilde{\mathcal{W}}_l^{(\rm Xe)}+\widetilde{\mathcal{W}}_l^{(\rm Ge)})
\label{eq:ratio1}
\end{align} 
and
\begin{align}
(\widetilde{\mathcal{W}}_l^{(\rm Xe)}-\widetilde{\mathcal{W}}_l^{(\rm Si)})/(\widetilde{\mathcal{W}}_l^{(\rm Xe)}+\widetilde{\mathcal{W}}_l^{(\rm Si)})
\label{eq:ratio2}
\end{align}
in the $(q,\Delta E)$ plane for $l=1$ and $l=2$.~Similarly, Fig.~\ref{fig:Wratio2} shows the same ratios now for $l=3$ and $l=4$.~In both cases, a superscript specifies the material for which $\widetilde{\mathcal{W}}_l(q,\Delta E)$ is evaluated.~Notice that only the regions above the black dashed lines are kinematically allowed in the two figures for our choice of $v_{\rm esc}$ and $v_e$.~For larger $v_{\rm esc}$, for example, smaller $|\vec q|$ would be accessible.~As expected, for $\Delta E$ below the xenon ionisation threshold, $\widetilde{\mathcal{W}}_l^{(\rm Ge)}$ and $\widetilde{\mathcal{W}}_l^{(\rm Si)}$ dominate over $\widetilde{\mathcal{W}}_l^{(\rm Xe)}$, showing that xenon and crystal detectors are mainly sensitive to complementary regions in the $(q,\Delta E)$ plane.~Despite this general trend, for deposited energies above about 30~eV the germanium $3d$ bands contribute significantly to $\widetilde{\mathcal{W}}_l^{(\rm Ge)}$ and the ratio $\widetilde{\mathcal{W}}_l^{(\rm Ge)}/\widetilde{\mathcal{W}}_l^{(\rm Xe)}$ can be of the order of 1 for kinematically allowed values of the momentum transfer.~While for $l=1$, $l=3$ and $l=4$ the response function $\bar{\mathcal{W}}_l(q,\Delta E)$ is always real, for $l=2$, we replace $\bar{\mathcal{W}}_2(q,\Delta E)$ with $|\bar{\mathcal{W}}_2(q,\Delta E)|$ in Eqs.~(\ref{eq:ratio1}) and (\ref{eq:ratio2}), as the former takes complex values for crystals.~In the case of isolated atoms, all material response functions are real.

With the notation of Eq.~(\ref{eq:Wcompact}), we can also write a general relation between products of material and DM response functions, and the modulus squared of the transition amplitude, namely
\begin{align}
\sum_{l=1}^r \Re\left[\mathcal{R}_l^*(q,v) \overline{\mathcal{W}}_l(q,\Delta E)\right] &=  \frac{2}{\pi}
\Delta E \sum_{\{1\},\{2\}} \int {\rm d}\Omega_q \, \overline{\overline{|\mathcal{M}_{1\rightarrow 2}(\vec v,\vec q)|^2}}
\,\delta(\Delta E - E_2 + E_1) \,,
\label{eq:WtoM}
\end{align}
where we used~\cite{Catena:2021qsr}
\begin{align}
\overline{\overline{|\mathcal{M}_{1\rightarrow 2}(\vec v,\vec q)|^2}}
&\equiv \frac{1}{2\pi} \int {\rm d} \phi \, {\overline{|\mathcal{M}_{1\rightarrow 2}(\vec v,\vec q)|^2}}_{\cos\theta=\xi} \nonumber\\
&= \sum_{l=1}^r \Re \left[ \mathcal{R}_l^*(q,v) \mathscr{B}_l(\vec q) \right] \,,
\label{eq:azimuthal}
\end{align}
$\theta$ and $\phi$ are polar and azimuthal angle of the velocity vector $\vec v$ in a frame with z-axis in the direction of $\vec q$,
\begin{align}
\xi= \frac{q}{2 m_{\rm DM} v} + \frac{\Delta E}{qv}\,,
\label{eq:xi}
\end{align}
and, finally, a single bar denotes a sum (average) over final (initial) DM/electron spin configurations.~In the evaluation of the azimuthal average in Eq.~(\ref{eq:azimuthal}), we use the following identifies~\cite{Catena:2021qsr}
\begin{align}
&\left. \frac{1}{2\pi}\int_0^{2\pi} \mathrm{d}\phi \, \vPerpEl\cdot\vec A\right|_{\cos\theta=\xi} =
\frac{\vec q}{m_e} \cdot \vec A \left(\frac{|\vec q|}{m_e}\right)^{-2} \frac{\vec q}{m_e} \cdot \vPerpEl \nonumber\\
&\left. \frac{1}{2\pi}\int_0^{2\pi} \mathrm{d}\phi \, \vPerpEl \cdot \left(\frac{\vec q}{m_e}\times\vec A\right) \right|_{\cos\theta=\xi} =0\,,
\label{eq:azimuthal_examples}
\end{align}
where $\vec A$ is an arbitrary three-dimensional vector.~We refer to~\cite{Catena:2021qsr} for a derivation of these equations.~In our numerical applications, we will be interested in the non-relativistic limit of Eq.~(\ref{eq:azimuthal}), 
\begin{align}
\overline{\overline{|\mathcal{M}_{1\rightarrow 2}(\vec v,\vec q)|^2}} &\simeq  \overline{\overline{| \mathcal{M}(\vec q,\vec v^\perp_{\rm el}) |^2_{\boldell=\mathbf{0}} |f_{1\rightarrow 2}(\vec q)|^2 }}
\nonumber\\[8pt]
&+2 m_e \Re \left[ \overline{\overline{\mathcal{M}(\vec q,\vec v^\perp_{\rm el}) _{\boldsymbol{\ell}=0}f_{1\rightarrow 2}(\vec q) [\nabla_{\boldsymbol{\ell}} \mathcal{M}^{*}(\vec q,\vec v^\perp_{\rm el}) ]_{\boldsymbol{\ell}=0}  \cdot \boldsymbol{f}^{*}_{1\rightarrow 2}(\vec q) }} \right] \nonumber\\[8pt]
&+ m_e^2 \overline{\overline{|\nabla_{\boldsymbol{\ell}} \mathcal{M}(\vec q,\vec v^\perp_{\rm el}) ]_{\boldsymbol{\ell}=0}  \cdot \boldsymbol{f}_{1\rightarrow 2}(\vec q) |^2}} \,.
\end{align}

\subsection{Tree-level application to specific models}
\label{sec:app}
In this section, we calculate the DM response functions $\mathcal{R}_l(q,v)$ associated with the models for DM-electron interactions of Sec.~\ref{sec:lagrangians}.~We refer to App.~\ref{app:details} for a detailed account of these calculations.
 
\begin{table}[t]
    \centering
    \begin{tabular*}{\columnwidth}{@{\extracolsep{\fill}}ll@{}}
    \toprule
        $\mathcal{O}_1 = \mathds{1}_{\rm DM}\mathds{1}_e$  & \vspace{0.15 cm} \\ 
        $\mathcal{O}_4 = \vec{S}_{\rm DM}\cdot \vec{S}_e$ & $\mathcal{O}_{11} = i\vec{S}_{\rm DM}\cdot\frac{\vec{q}}{m_e}\mathds{1}_e$ \\ 
        $\mathcal{O}_5 = i\vec{S}_{\rm DM}\cdot\left(\frac{\vec{q}}{m_e}\times\vt\right)\mathds{1}_e$ &  $\mathcal{O}_{14} = i\left(\vec{S}_{\rm DM}\cdot \frac{\vec{q}}{m_e}\right)\left(\vec{S}_e\cdot \vt\right)$ \\
        $\mathcal{O}_6 = \left(\vec{S}_{\rm DM}\cdot\frac{\vec{q}}{m_e}\right) \left(\vec{S}_e\cdot\frac{\vec{q}}{m_e}\right)$ &$\mathcal{O}_{15} = -\left(\vec{S}_{\rm DM}\cdot \frac{\vec{q}}{m_e}\right)\left[ \left(\vec{S}_e\times \vt \right) \cdot \frac{\vec{q}}{m_e}\right] $  \\
        $\mathcal{O}_7 = \vec{S}_e\cdot \vt\mathds{1}_{\rm DM}$ & $\mathcal{O}_{17}=i \frac{\vec{q}}{m_e} \cdot \boldsymbol{\mathcal{S}} \cdot \vt \mathds{1}_e$  \\
        $\mathcal{O}_8 = \vec{S}_{\rm DM}\cdot \vt\mathds{1}_e$ & $\mathcal{O}_{18}=i \frac{\vec{q}}{m_e} \cdot \boldsymbol{\mathcal{S}}  \cdot \vec{S}_e$  \\
        $\mathcal{O}_9 = i\vec{S}_{\rm DM}\cdot\left(\vec{S}_e\times\frac{\vec{q}}{m_e}\right)$ & $\mathcal{O}_{19} = \frac{\vec{q}}{m_e} \cdot \boldsymbol{\mathcal{S}} \cdot \frac{\vec{q}}{m_e}\mathds{1}_e$  \\
       $\mathcal{O}_{10} = i\vec{S}_e\cdot\frac{\vec{q}}{m_e}\mathds{1}_{\rm DM}$   & $\mathcal{O}_{20} = \left(\vec{S}_e \times \frac{\vec{q}}{m_e} \right) \cdot \boldsymbol{\mathcal{S}} \cdot \frac{\vec{q}}{m_e}$  \\
    \bottomrule
    \end{tabular*}
    \caption{List of operators that can contribute to the transition amplitude for the models considered here.~We denote by $\mathds{1}_{\rm DM}$ ($\mathds{1}_e$) the identity in the DM (electron) spin space.~Similarly, $\vec{S}_{\rm DM}$ ($\vec{S}_e$) is the DM (electron) spin operator.~For spin-0 DM, $\langle \vec{S}_{\rm DM} \rangle=0$.~For spin-1/2 DM $\langle \vec{S}_{\rm DM} \rangle=\delta^{r' r} \xi^{s'}\boldsymbol{\sigma} \xi^s/2$, where $r$ ($s$) labels the initial electron (DM) spin configuration, $r'$ ($s'$) identifies the final electron (DM) spin configuration and $\boldsymbol=(\sigma_1,\sigma_2,\sigma_3)$ consists of the three Pauli matrices.~For spin-1 DM, $\langle \vec{S}_{\rm DM} \rangle = -i \delta^{r' r} \vec e'_{s'}\times \vec e_{s}$, where $\vec e_s$ and $\vec e'_{s'}$ are three-dimensional polarisation vectors.~Finally, $\langle \boldsymbol{\mathcal{S}}  \rangle = \delta^{r' r}(e_{si} e'_{s'j} + e_{sj} e'_{s'i})/2$.~See App.~\ref{app:identities} for further details.}
\label{tab:operators}
\end{table} 
  
\subsubsection{Scalar dark matter}
In the case of  {\it scalar DM with a scalar mediator}, the free amplitude for DM-electron scattering is
\begin{align}
i\mathcal{M} = i g_1 m_S \left[ \frac{i}{(k'-k)^2-m_\phi^2} \right] \left[ \bar{u}^{r'}(k') i ( h_1 + ih_2 \gamma^5 ) u^r(k) \right] \,,
\label{eq:M00rel}
\end{align}
where $u^{r}(k)$, $r=1,2$ and $\bar{u}^{r'}(k')$, $r'=1,2$ are free Dirac spinors for the initial and final state electron, while $k$ and $k'$ are four-momenta.~In the non-relativistic limit, Eq.~(\ref{eq:M00rel}) can be written as
\begin{align}
i\mathcal{M} \simeq i 2m_S m_e \left( \frac{g_1h_1}{|\vec q|^2+m_\phi^2} \langle \hat{\mathcal{O}}_1\rangle - \frac{g_1 h_2}{|\vec q|^2+m_\phi^2} \langle \hat{\mathcal{O}}_{10}\rangle\right)  \,,
\label{eq:M00}
\end{align}
where the operators $\hat{\mathcal{O}}_1$ and $\hat{\mathcal{O}}_{10}$ are given in Tab.~\ref{tab:operators}, and angle brackets  in general denote an expectation value between initial, $s/r$, and final,  $s'/r'$, DM/electron spin configurations.~For example, $\langle \hat{\mathcal{O}}_1\rangle =\delta^{r'r}\delta^{s's}$.~For scalar DM $\delta^{s's}\rightarrow 1$, while $s,s'=1,2$ ($s,s'=1,2,3$) for spin-1/2 (spin-1) DM.~By inserting Eq.~(\ref{eq:M00}) into Eq.~(\ref{eq:M_expansion}), we find the modulus squared transition amplitude 
\begin{align}
\overline{|M_{1\rightarrow 2}|^2} = \left( c_1^2 + \frac{1}{4} c_{10}^2 \frac{|\vec q|^2}{m_e^2} \right) \mathscr{B}_1\,,
\label{eq:M:01_squared}
\end{align}
where the coupling constants $c_1$ and $c_{10}$ are related to the coupling constants and masses of the underlying model as in Tab.~\ref{tab:S0coeff}.~Notice the $1/\Gamma(u,t)$ factor in the definition of $c_1$ and $c_{10}$, where
\begin{align}
\Gamma(u,t)^{-1}\equiv 4t (1+u^2)^{-1}\,,
\end{align} 
arising from the mediator propagator.~Comparing Eq.~(\ref{eq:M:01_squared}) with the second line in Eq.~(\ref{eq:azimuthal}), we find
\begin{align}
\mathcal{R}_1(q,v)= \left( c_1^2 + \frac{1}{4} c_{10}^2 \frac{|\vec q|^2}{m_e^2} \right)\,,
\label{eq:R00}
\end{align}
while $\mathcal{R}_2(q,v)=\mathcal{R}_3(q,v)=\mathcal{R}_4(q,v)=\mathcal{R}_5(q,v)=0$.~The case of spin 0 DM with a scalar mediator is thus not characterised by new features, since the observable electron transition rate can in this case be computed by using the response function $\overline{\mathcal{W}}_1$, as in the dark photon model. 

As a first non trivial example where the ionisation and crystal form factors commonly used in the case of the dark photon model provide a poor description of the rate of electron transitions induced by DM scattering, let us consider the case of  {\it scalar DM with a vector mediator}.~For this model, the free amplitude for DM-electron scattering can be written as follows
\begin{align}
i\mathcal{M}=ig_4 P_\mu \left[ \frac{-ig^{\mu\nu}}{(k'-k)^2-m_G^2} \right] \left[ \bar{u}^{r'}(k') i ( h_3 \gamma_\nu+ ih_4 \gamma_\nu\gamma^5 ) u^r(k) \right] \,,
\end{align}
where $P=p+p'$, and $p$ ($p'$) is the initial (final) DM four-momentum.~In the non-relativistic limit,
\begin{align}
i\mathcal{M} \simeq i2m_S m_e \left(  -\frac{2 g_4 h_3}{|\vec q|^2+m_G^2} \langle \hat{\mathcal{O}}_{1} \rangle+  \frac{4g_4 h_4}{|\vec q|^2+m_G^2} \langle \hat{\mathcal{O}}_{7} \rangle \right) \,,
\end{align}
where the operators $\hat{\mathcal{O}}_1$ and $\hat{\mathcal{O}}_{7}$ are listed in Tab.~\ref{tab:operators}.~For the azimuthal average of the squared modulus of the transition amplitude $\mathcal{M}_{1\rightarrow 2}$ (defined in Eq.~(\ref{eq:azimuthal}) and used in Eq.~(\ref{eq:WtoM})), we now find
\begin{align}
\overline{\overline{|\mathcal{M}_{1\rightarrow 2}(\vec v,\vec q)|^2}} = \left( c_1^2 + \frac{1}{4} c_7^2  \,|\vec v^\perp_{\rm el}|^2\right) \mathscr{B}_1
- \frac{1}{2} c_7^2\, \left( \frac{\vec q}{m_e} \cdot \vec v^\perp_{\rm el} \right) \frac{m_e^2}{|\vec q|^2} \,\mathscr{B}_2
+ \frac{1}{4} c_7^2 \, \mathscr{B}_3
\,,
\end{align}
which implies
\begin{align}
\mathcal{R}_1(q,v)&= \left( c_1^2 + \frac{1}{4} c_7^2 \,|\vec v^\perp_{\rm el}|^2\right) \nonumber\\
\mathcal{R}_2(q,v)&= - \frac{1}{2} c_7^2 \left( \frac{\vec q}{m_e} \cdot \vec v^\perp_{\rm el} \right) \frac{m_e^2}{|\vec q|^2}  \nonumber\\
\mathcal{R}_3(q,v)&= \frac{1}{4} c_7^2\,,
\label{eq:R01}
\end{align}
with $\mathcal{R}_4(q,v)=\mathcal{R}_5(q,v)=0$.~Tab.~\ref{tab:S0coeff} relates $c_1$ and $c_7$ to the parameters of the underling DM model.~As in the previous example, in Tab.~\ref{tab:S0coeff} we include a $1/\Gamma(u,t)$ factor in the definition of $c_1$ and $c_7$.

 \begin{table}[t]
    \centering
    \begin{tabular}{l l l l l l l || l l l l l l l}
    \toprule
        \multicolumn{7}{l||}{{\rm Scalar Mediator}, } 
                 & \multicolumn{7}{l}{{\rm Vector Mediator},} \\
                    \multicolumn{7}{l||}{$u=|\vec q|/m_\phi$, $t=m_S m_e/m^2_\phi$} 
                    & \multicolumn{7}{l}{$u=|\vec q|/m_G$, $t=m_S m_e/m^2_G$} \\
                  \hline
                 & & & & & & & & & & & \\
         \(\displaystyle c_1 = \frac{1}{2}\frac{g_1h_1}{\Gamma(u,t)}\) 
            & & & & &\(\displaystyle c_{10} = -\frac{1}{2}\frac{g_1h_2}{\Gamma(u,t)}\) 
             & 
                & \(\displaystyle c_1 = -\frac{g_4h_3}{\Gamma(u,t)}\)
                    & & & && \(\displaystyle c_7 = \frac{2g_4h_4}{\Gamma(u,t)}\)
                     &   \\
     \bottomrule
    \end{tabular}
    \caption{Relation between the coupling constants in the Lagrangians of Eq.~(\ref{eq:Lss}) and Eq.~(\ref{eq:Lsv}), and the non-relativistic coupling constants in the DM response functions for, respectively, scalar DM with a scalar mediator, Eq.~(\ref{eq:R00}), and scalar DM with a vector mediator, Eq.~(\ref{eq:R01}).~The dimensionless function $\Gamma(u,t)^{-1}\equiv 4t (1+u^2)^{-1}$ accounts for the dependence of the non-relativistic coupling constants on the momentum transfer $\vec q$ and the mediator mass, $m_\phi$ or $m_G$.}
    \label{tab:S0coeff}
\end{table}

\subsubsection{Fermionic dark matter}
The calculation of the DM response functions $\mathcal{R}_{l}(q,v)$ for fermionic DM proceeds as in the case of spin-0 DM, which we discussed in detail in the previous subsection.~We provide the details of such a calculation in App.~\ref{app:details} and App.~\ref{app:identities}.~Here, we limit ourselves to listing the final results, placing the emphasis on models that predict material responses different from the standard ionisation and crystal form factors.~For {\it fermionic DM with a scalar mediator}, we find
\begin{align}
   \mathcal{R}_1(q,v) = c_1^2 + \frac{1}{16}c_6^2\frac{|\vec q|^4}{m_e^4} + \frac{1}{4}c_{10}^2\frac{|\vec q|^2}{m_e^2} + \frac{1}{4}c_{11}^2\frac{|\vec q|^2}{m_e^2} \,,
   \label{eq:R1/20}
\end{align}
while $\mathcal{R}_2(q,v)=\mathcal{R}_3(q,v)=\mathcal{R}_4(q,v)=\mathcal{R}_5(q,v)=0$.~As in the case of spin-0 DM with a scalar mediator (see Eq.~(\ref{eq:R00})), we find that in a scenario where DM is made of spin-1/2 particles interacting with electrons through the exchange of a scalar mediator particle the standard ionisation and crystal form factors can properly describe the response of materials to an external DM perturbation.~In contrast, for {\it fermionic DM with a vector mediator}, we find 
\begin{align}
\mathcal{R}_1(q,v) &= c_1^2 + \frac{3}{16}c_4^2 + \frac{1}{4}c_7^2|\vec v^\perp_{\rm el}|^2 + \frac{1}{4} c_8^2 |\vec v^\perp_{\rm el}|^2 + \frac{1}{8}c_9^2\frac{|\vec q|^2}{m_e^2} \nonumber\\
\mathcal{R}_2(q,v) &= -\left(\frac{\vec q}{m_e}\cdot\vec v^\perp_{\rm el}\right) \frac{m_e^2}{|\vec q|^2}\left(\frac{1}{2}c_7^2 + \frac{1}{2}c_8^2\right) \nonumber \\
\mathcal{R}_3(q,v) &= \frac{1}{4}c_7^2 + \frac{1}{4}c_8^2 \,,
 \label{eq:R1/21}
\end{align}
$\mathcal{R}_4(q,v)=\mathcal{R}_5(q,v)=0$.~We therefore conclude that it is crucial to evaluate the material response functions $\bar{\mathcal{W}}_2(q,\Delta E)$ and $\bar{\mathcal{W}}_3(q,\Delta E)$, in addition to the standard ionisation and crystal form factors, to consistently compare theoretical predictions with experimental data in a model with fermionic DM and a vector mediator.~For example, neglecting the $\bar{\mathcal{W}}_2(q,\Delta E)$ and $\bar{\mathcal{W}}_3(q,\Delta E)$ response functions when computing an exclusion limit from a null result, or when interpreting a signal at a DM direct detection experiment, would lead to strongly biased results if nature is indeed described by fermionic DM with a vector mediator.

\begin{table}[t]
    \centering
       \begin{tabular}{l l l l || l l l l l}
  	\toprule
        \multicolumn{4}{l||}{{\rm Scalar Mediator},}
                & \multicolumn{5}{l}{{\rm Vector Mediator},} \\
                \multicolumn{4}{l||}{$u=|\vec q|/m_\phi$, $t=m_\chi m_e/m^2_\phi$}
                & \multicolumn{5}{l}{$u=|\vec q|/m_G$, $t=m_\chi m_e/m^2_G$  \phantom{xxxxxxxxxxxxxxx}} \\
                \hline
                & & & &  &  & &\\
        \(\displaystyle c_1 = \frac{\lambda_1h_1}{\Gamma(u,t)} \)
            & & &   \(\displaystyle c_6 = \frac{\lambda_2h_2}{\Gamma(u,t)} \frac{m_e}{m_\chi}\) 
                &\(\displaystyle c_1 = -\frac{\lambda_3h_3}{\Gamma(u,t)} \)
                    & & &  &\(\displaystyle c_4\ = \frac{4\lambda_4h_4}{\Gamma(u,t)}\)
                    \\
        \(\displaystyle c_{10} = -\frac{\lambda_1h_2}{\Gamma(u,t)}\)
            & & &   \(\displaystyle c_{11} = \frac{\lambda_2h_1}{\Gamma(u,t)} \frac{m_e}{m_\chi}\)
                &\(\displaystyle c_7 = \frac{2\lambda_3h_4}{\Gamma(u,t)}\)
                   & & &  & \(\displaystyle c_8 = -\frac{2\lambda_4h_3}{\Gamma(u,t)}\)   \\ 
            & & & & \multicolumn{2}{l}{\(\displaystyle c_9 = \frac{2\lambda_3h_4}{\Gamma(u,t)} \frac{m_e}{m_\chi} + \frac{2\lambda_4h_3}{\Gamma(u,t)}\)}
            &  \\
  	\bottomrule
    \end{tabular}
     \caption{Relation between the coupling constants in the Lagrangians of Eq.~(\ref{eq:Lfs}) and Eq.~(\ref{eq:Lfv}) and the non-relativistic coupling constants in the DM response functions for, respectively, fermionic DM with a scalar mediator, Eq.~(\ref{eq:R1/20}), and fermionic DM with a vector mediator, Eq.~(\ref{eq:R1/21}).~Here, we introduced the dimensionless function $\Gamma(u,t)^{-1}\equiv 4t (1+u^2)^{-1}$ to account for the dependence of the non-relativistic coupling constants on $\vec q$ and the mediator mass.}
    \label{tab:S0.5coeff}
\end{table}

\subsubsection{Vector dark matter}
In the case of {\it spin-1 DM with a scalar mediator} we find that just a single response function contributes to the observed rate of electron transitions induced by DM scattering in a material, namely
\begin{align}
\mathcal{R}_1(q,v)= c_1^2 + \frac{1}{4}c_{10}^2\frac{|\vec q|^2}{m_e^2} \,,
 \label{eq:R10}
\end{align}
while $\mathcal{R}_2(q,v)=\mathcal{R}_3(q,v)=\mathcal{R}_4(q,v)=\mathcal{R}_5(q,v)=0$.~We therefore find that in all models where the interaction between DM and the electrons bound to a material is mediated by the exchange of a scalar mediator the rate of DM-induced electron transitions can be computed by using the standard ionisation and crystal form factors.~Turning this argument around, we find that experimental evidence for contributions to the theoretical rate $\mathscr{R}_{\rm theory}$ from $\mathcal{W}_l$, with $l\neq 1$, would point towards DM-electron interactions mediated by particles with spin different from zero.

\begin{table}[t]
    \centering
    \caption{The same as in Tab.~\ref{tab:S0coeff} and Tab.~\ref{tab:S0.5coeff}, now for vector DM.}
    \begin{tabular}{l l}
    \toprule
    \multicolumn{2}{c}{\vspace{3pt}Scalar Mediator, $u=|\vec q|/m_\phi$, $t=m_Xm_e/m^2_\phi$} \\ 
    \hline
    & \\
    \(\displaystyle c_1 = \frac{1}{2} \frac{b_1h_1}{\Gamma(u,t)} \)  & 
    \(\displaystyle c_{10} = - \frac{1}{2}\frac{b_1h_2}{\Gamma(u,t)} \)\vspace{3pt}\\
    \toprule
     \multicolumn{2}{c}{Vector Mediator, $u=|\vec q|/m_G$, $t=m_Xm_e/m^2_G$} \\
      \hline
    & \\
    \(\displaystyle c_1 = -\frac{b_5h_3}{\Gamma(u,t)}\)
         & \\ \vspace{3pt}
    \(\displaystyle c_4 = \frac{\Im(b_6)h_3}{\Gamma(u,t)} \frac{m_e}{2m_X}\frac{|\vec q\,|^2}{m_e^2} - 
            \frac{2\Re(b_7)h_4}{\Gamma(u,t)} \phantom{xxxxxxx}\)
                & \(\displaystyle c_{11} = \frac{\Im(b_7)h_3}{\Gamma(u,t)} \frac{m_e}{2m_X}\)\\ \vspace{3pt}
    \(\displaystyle c_5 = \frac{\Im(b_6)h_3}{\Gamma(u,t)} \frac{m_e}{2m_X}\)
                &\(\displaystyle c_{14} = -\frac{\Im(b_7)h_4}{\Gamma(u,t)} \frac{m_e}{m_X}\)\\ \vspace{3pt}
    \(\displaystyle c_6 = -\frac{\Im(b_6)h_3}{\Gamma(u,t)} \frac{m_e}{2m_X}\)
                &\(\displaystyle c_{17} = -\frac{\Re(b_6)h_3}{\Gamma(u,t)} \frac{m_e}{m_X}\)\\ \vspace{3pt}
    \(\displaystyle c_7 = \frac{2b_5h_4}{\Gamma(u,t)}\)
                &\(\displaystyle c_{18} = \frac{2\Re(b_6)h_4}{\Gamma(u,t)} \frac{m_e}{m_X}\)\\ \vspace{3pt}
    \(\displaystyle c_8 = \frac{\Re(b_7)h_3}{\Gamma(u,t)}\)
                &\(\displaystyle c_{19} = \frac{\Im(b_6)h_3}{\Gamma(u,t)} \frac{m_e^2}{2m_X^2}\)\\ \vspace{3pt}
    \(\displaystyle c_9 = \frac{\Im(b_6)h_4}{\Gamma(u,t)} \frac{m_e}{m_X} - 
                \frac{\Re(b_7)h_3}{\Gamma(u,t)}\)
                &\(\displaystyle c_{20} = -\frac{\Re(b_6)h_3}{\Gamma(u,t)} \frac{m_e}{m_X} - 
                            \frac{\Im(b_7)h_4}{\Gamma(u,t)} \frac{m_e^2}{m_X^2}\)\\ 
    \bottomrule
    \end{tabular}
    \label{tab:S1coeff}
\end{table}

The case of {\it spin-1 DM with a vector mediator} exhibits a richer phenomenology, with several material response functions possibly contributing to the theoretical rate $\mathscr{R}_{\rm theory}$.~Specifically, 
\begin{align}
    \mathcal{R}_1(q,v)
    &=c_1^2 + 
    \frac{1}{2}c_4^2 + 
    \frac{2}{3}c_5^2\left|\frqm\times\vvp\right|^2 + 
    \frac{1}{6}c_6^2\frac{\qAbs^4}{m_e^4} +
    \frac{1}{4}c_7^2\vAbs^2 +
    \frac{2}{3}c_8^2\vAbs^2 \nonumber\\
    & \qquad \,+ 
    \frac{1}{3}c_9^2\frac{\qAbs^2}{m_e^2} + 
    \frac{2}{3}c_{11}^2\frac{\qAbs^2}{m_e^2} + 
    \frac{1}{6}c_{14}^2\vAbs^2\frac{\qAbs^2}{m_e^2} \nonumber\\
    & \qquad \,+ 
    \frac{1}{6}c_{17}^2\left(\left|\frqm\cdot\vvp\right|^2 + \frac{\qAbs^2}{m_e^2}\vAbs^2\right) + 
    \frac{1}{6}c_{18}^2\frac{\qAbs^2}{m_e^2} + 
    \frac{1}{3}c_{19}^2\frac{\qAbs^4}{m_e^4} \nonumber \\
    & \qquad \,+ 
    \frac{1}{12}c_{20}^2\frac{\qAbs^4}{m_e^4} + 
    \frac{1}{3}c_4c_6\frac{\qAbs^2}{m_e^2} +\frac{2}{3} c_1c_{19}\frac{\qAbs^2}{m_e^2} \nonumber\\
\Re[\mathcal{R}_2(v,q)] &= -\left( \frqm\cdot\vvp \right)\left[\frac{m_e^2}{\qAbs^2}
    \left(\frac{1}{2}c_7^2 + \frac{4}{3}c_8^2\right)
    + \frac{1}{3}c_{14}^2 + \frac{2}{3}c_{17}^2 \right] \nonumber\\
\Im[\mathcal{R}_2(v,q)] &= \frac{1}{2} c_7 c_{10} -\frac{1}{3}c_4 c_{14} + \frac{4}{3} c_8 c_{11} - \frac{1}{3} c_6 c_{14}  \frac{\qAbs^2}{m_e^2} \nonumber\\    
\mathcal{R}_3(v,q) &= \frac{1}{4}c_7^2 + \frac{2}{3}c_8^2 + \frac{\qAbs^2}{m_e^2}\left(\frac{2}{3} c_5^2 + \frac{1}{6}c_{14}^2 + \frac{1}{6}c_{17}^2\right) \nonumber\\
\mathcal{R}_4(v,q) &= -\frac{2}{3} c_5^2 + \frac{1}{6}c_{17}^2 \,,
\label{eq:R11}
\end{align}
with $R_5(q,v)=0$.~We refer to App.~\ref{app:details} and App.~\ref{app:identities} for all details underling the derivation of Eq.~(\ref{eq:R11}).~Interestingly, none of the models considered here can generate the response function $\bar{\mathcal{W}}_5$.

\begin{figure}[t]
\centering	
\subfloat{	
\begin{fmffile}{DM00}
 \setlength{\unitlength}{.5mm}\large
\begin{fmfgraph*}(70,80)
\fmftop{t1,t2}
\fmfbottom{b1,b2}
\fmf{fermion}{t1,v1}
\fmf{fermion}{v2,t2}
\fmf{dots,label=$\phi \ $, label.side=right, tension=.5}{v1,v3}
\fmf{dots,label=$\ \phi$,tension=.5}{v3,v2}
\fmf{fermion,label=$e^{-}$,tension=.5}{v1,v2}
\fmf{dashes}{b1,v3,b2}
\fmflabel{$e^{-}$}{t1}
\fmflabel{$e^{-}$}{t2}
\fmflabel{$S$}{b1}
\fmflabel{$S$}{b2}
\end{fmfgraph*}
\end{fmffile}
}
\hspace{2 cm}
\subfloat{
\begin{fmffile}{DM01}
 \setlength{\unitlength}{.5mm}\large
\begin{fmfgraph*}(70,80)
\fmftop{t1,t2}
\fmfbottom{b1,b2}
\fmf{fermion}{t1,v1}
\fmf{fermion}{v2,t2}
\fmf{dbl_curly,label=$G \ \ $, label.side=right, tension=.5}{v1,v3}
\fmf{dbl_curly,label=$\ \ G$,tension=.5}{v3,v2}
\fmf{fermion,label=$e^{-}$,tension=.5}{v1,v2}
\fmf{dashes}{b1,v3,b2}
\fmflabel{$e^{-}$}{t1}
\fmflabel{$e^{-}$}{t2}
\fmflabel{$S$}{b1}
\fmflabel{$S$}{b2}
\end{fmfgraph*}
\end{fmffile}
}

\vspace{0.5 cm}
\subfloat{
\begin{fmffile}{DM10}
 \setlength{\unitlength}{.5mm}\large
\begin{fmfgraph*}(70,80)
\fmftop{t1,t2}
\fmfbottom{b1,b2}
\fmf{fermion}{t1,v1}
\fmf{fermion}{v2,t2}
\fmf{dots,label=$\phi \ $, label.side=right, tension=.5}{v1,v3}
\fmf{dots,label=$\ \phi$,tension=.5}{v3,v2}
\fmf{fermion,label=$e^{-}$,tension=.5}{v1,v2}
\fmf{curly}{b1,v3,b2}
\fmflabel{$e^{-}$}{t1}
\fmflabel{$e^{-}$}{t2}
\fmflabel{$X$}{b1}
\fmflabel{$X$}{b2}
\end{fmfgraph*}
\end{fmffile}
}
\hspace{2 cm}
\subfloat{
\begin{fmffile}{DM11}
 \setlength{\unitlength}{.5mm}\large
\begin{fmfgraph*}(70,80)
\fmftop{t1,t2}
\fmfbottom{b1,b2}
\fmf{fermion}{t1,v1}
\fmf{fermion}{v2,t2}
\fmf{dbl_curly,label=$G \ \ $,label.side=right,tension=.5}{v1,v3}
\fmf{dbl_curly,label=$\ \ G$,tension=.5}{v3,v2}
\fmf{fermion,label=$e^{-}$,tension=.5}{v1,v2}
\fmf{curly}{b1,v3,b2}
\fmflabel{$e^{-}$}{t1}
\fmflabel{$e^{-}$}{t2}
\fmflabel{$X$}{b1}
\fmflabel{$X$}{b2}
\end{fmfgraph*}
\end{fmffile}
}
\caption{One-loop Feynman diagrams for DM-electron scattering assuming:~1) spin-0 DM with a scalar mediator (top left), 2) spin-0 DM with a vector mediator (top right), 3) spin-1 DM with a scalar mediator (bottom left) and, finally, 4) spin-1 DM with a vector mediator (bottom right).~The particles in the processes are denoted by the letters used  for the associated fields.}
\label{fig:loops}
\end{figure}
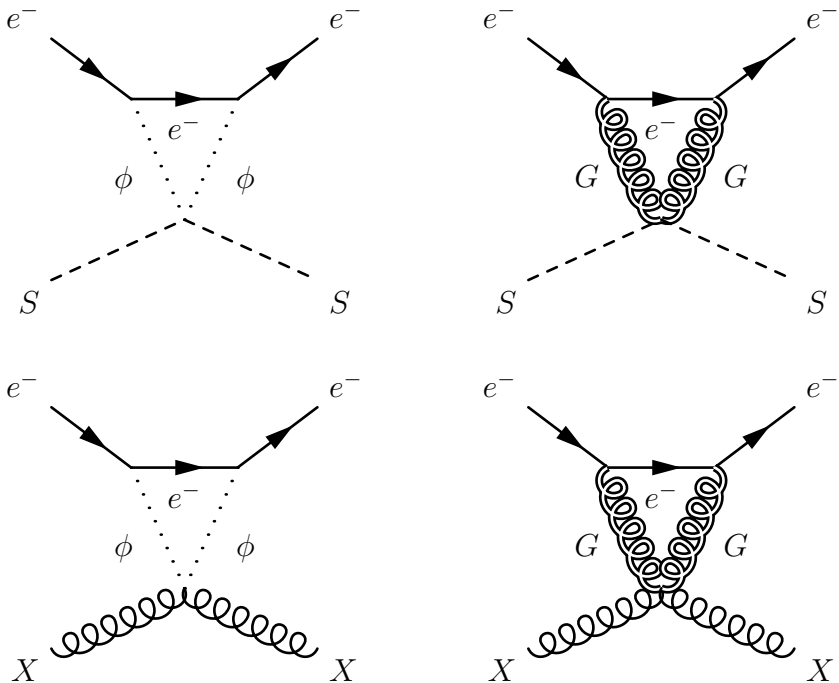

\subsection{One-loop application to specific models}
\label{sec:loop}
So far, we focused on DM-electron interactions arising at tree level.~These required a cubic DM-DM-mediator interaction vertex in the Lagragian.~It is interesting to note that even when such cubic vertices are absent at tree level, DM-electron interactions can be generated at the one-loop level via a quartic DM-DM-mediator-mediator vertex, as shown in Fig.~\ref{fig:loops} in four specific cases:~1) scalar DM with a scalar mediator, 2) scalar DM with a vector mediator, 3) vector DM with a scalar mediator, and, finally 4) vector DM with a vector mediator.~In App.~\ref{app:loops}, we list the free electron amplitude for the four diagrams in Fig.~\ref{fig:loops}.~They are all proportional to $\NRop{1}$, and therefore contribute to the observable rate of electron transitions induced by DM scattering in a detector material, $\mathscr{R}_{\rm theory}$, via the response function $\bar{\mathcal{W}}_1(q,\Delta E)$ only.~From the point of view of material physics and for the purposes of our work, this family of DM-electron interaction models is therefore equivalent to the dark photon model.~For this reason, here we do not investigate the effects of loop-induced DM-electron interactions any further.

\section{Comparison with experimental data}
\label{sec:results}
In this section, we compare the rate of electron transitions predicted from the models of Sec.~\ref{sec:lagrangians} with experimental data released by the XENON10~\cite{Angle:2011th}, XENON1T\cite{Aprile:2019xxb}, EDELWEISS~\cite{EDELWEISS:2020fxc} and SENSEI~\cite{SENSEI:2020dpa} collaborations.~As we will see below, for xenon experiments we adopt the data analysis strategies we used in~\cite{Catena:2019gfa}.~For the analysis of results from crystal detectors, we rely on~\cite{Catena:2021qsr}.

\begin{table}[t]
	\centering
	\begin{tabular}{|ll|c|ll|}
	\cline{1-2}\cline{4-5}
	\textbf{XENON10\phantom{xxxxxx}}			&			&\phantom{space}&\textbf{XENON1T\phantom{xxxxxx}}	&			\\
	bin~[S2]				&observed events		&&bin~[S2]				&observed events		\\
	\cline{1-2}\cline{4-5}
	$\text{[}$14,41)			&126			&&$\text{[}$150,200)		&8			\\
	$\text{[}$41,68)			&60			&&$\text{[}$200,250)		&7			\\
	$\text{[}$68,95)			&12			&&$\text{[}$250,300)	&2			\\
	$\text{[}$95,122)		&3			&&$\text{[}$300,350)	&1			\\
	$\text{[}$122,149)		&2			&&-	&-			\\
	$\text{[}$149,176)		&0			&&-	&-			\\
	$\text{[}$176,203)		&2			&&-	&-			\\
	\cline{1-2}\cline{4-5}
	\end{tabular}
	\caption{Events in selected {\rm S2} bins observed by XENON10~\cite{Angle:2011th} (left) and XENON1T~\cite{Aprile:2019xxb} (right).}
	\label{tab:S2events}
\end{table}

\subsection{Direct detection data}
The XENON10 and XENON1T experiments are dual phase time projection chambers (TPCs) filled with liquid and gaseous xenon.~They have been searching for sub-GeV DM particles in events where xenon atoms are ionised by DM-electron scattering in the liquid xenon detector component.~In each such event, primary electrons from the ionised atoms are expected to propagate in the liquid xenon, producing a number $n_e$ of secondary electrons.~These secondary electrons are then extracted from the liquid target by means of an external electric field, and drifted into the gaseous xenon detector component, where they eventually generate a scintillation signal.~This signal is finally recorded by photomultiplier tubes through the observation of ${\rm S2}$ associated photoelectrons~\cite{Angle:2011th,Aprile:2019xxb}.~Since ${\rm S2}$ is the actually observed quantity in xenon TPCs experiments, we need to rewrite our theoretical rate, Eq.~(\ref{eq:Robs}), as a function of ${\rm S2}$.

Let us denote by $\mathscr{P}(n_e|E_e)$ the probability density to produce $n_e$ secondary electrons in a DM-electron scattering event when the energy of the primary electron is $E_e=k^{\prime2}/(2 m_e)$.~Furthermore, let $\mathscr{P}({\rm S2}|n_e)$ be the probability density to observe ${\rm S2}$ photoelectrons when $n_e$ electrons drifted from the liquid xenon target into the gaseous phase.~Then, the rate of DM-electron scattering events with a number of observed photoelectrons between ${\rm S2}$ and ${\rm S2}+{\rm d}{\rm S2}$ is~\cite{Catena:2019gfa} 
\begin{align}
{\rm d }\mathscr{R}_{\rm obs} &={\rm d} {\rm S2}\,\epsilon({\rm S2}) \frac{n_{\rm DM} 
}{128\pi m_{\rm DM}^2 m_e^2}\int \mathrm{d} (\ln\Delta E)\int \mathrm{d} q \, q \,\widehat{\eta}\left(q, \Delta E
\right) \sum_{n_e=1}^{\infty}  \mathscr{P}({\rm S2}|n_e)
\nonumber\\
&\times \sum_{l=1}^r \Re\left[\mathcal{R}_l^*(q,v) \overline{\mathcal{W}}_l(q,\Delta E;n_e)\right]\,,
\label{eq:Ratoms_obs}
\end{align}
where $\epsilon({\rm S2})$ is the experiment-specific detection efficiency, and we introduced the modified response
\begin{align}
    \overline{\mathcal{W}}_l(q,\Delta E;n_e)\equiv&\int \mathrm{d}\Omega_q\, \mathcal{W}_l(\vec q,\Delta E;n_e) \,,
\end{align}
with
\begin{align}
    \mathcal{W}_l(\vec q,\Delta E;n_e)&=\frac{2}{\pi} \Delta E  \sum_{n,\ell,m} \sum_{\ell',m'}  
   \int \frac{V k^{\prime 2} \,\mathrm{d}k^\prime}{(2\pi)^3} \,\mathscr{B}_l \,   \mathscr{P}(n_e|E_e)\,
   \delta(\Delta E - E_{n\ell m} + E_{k'\ell' m'})
    \label{eq:Watoms2}\,.
\end{align}
It is important to note that Eq.~(\ref{eq:Ratoms_obs}) gives the rate of observable events per atom in the detector, not the total rate.~The total rate of observable events per unit detector mass is found by multiplying Eq.~(\ref{eq:Ratoms_obs}) by the number of xenon atoms in the detector, $n_{\rm Xe}$, and then dividing by the detector mass, $M_{\rm det}= n_{\rm Xe} m_{\rm Xe}$, where $m_{\rm Xe}$ is the mass of a xenon atom.~One finds,
\begin{align}
{\rm d }\widetilde{\mathscr{R}}_{\rm atoms} \equiv {\rm d }\mathscr{R}_{\rm obs}/m_{\rm Xe} \,.
\label{eq:Ratoms_obs2}
\end{align}
We refer to~\cite{Catena:2019gfa} for further details on $\mathscr{P}(n_e|E_e)$, $\mathscr{P}({\rm S2}|n_e)$ and $\epsilon({\rm S2})$.~By integrating Eq.~(\ref{eq:Ratoms_obs2}) over the ${\rm S2}$ bins in Tab.~\ref{tab:S2events}, and multiplying the result by the experimental exposure (15 kg~days for XENON10 and 80755 kg~days for XENON1T), we obtain the expected number of DM signal events in the given ${\rm S2}$ bin.~We can then use the statistical methods reviewed below to compare our predictions with the XENON10 \cite{Angle:2011th} and XENON1T \cite{Aprile:2019xxb} results we report in Tab.~\ref{tab:S2events}.

\begin{table}[t]
	\centering
	\begin{tabular}{|lll|c|lll|}
	\cline{1-3}\cline{5-7}
	\textbf{EDELWEISS}			&				&&\phantom{xx}&\textbf{SENSEI}	&&			\\
	$[Q]$					&obs. events	&	exp. [g-day] && $[Q]$				&obs. events	& exp.~[g-day] 	\\
	\cline{1-3}\cline{5-7}
	1			&5814		& 3.23 & &1	&758		&1.38	\\						
	2			&44706		& 17.76 & &2	&5		&2.09	\\						
	3			&2718		& 80.72	 & &3	&0		&9.03	\\						
	4			&227			& 80.72	 & &4	&0		&9.10	\\						
	\cline{1-3}\cline{5-7}
	\end{tabular}
	\caption{Events corresponding to $Q=1,2,3,4$ observed by the EDELWEISS~\cite{EDELWEISS:2020fxc} and SENSEI~\cite{SENSEI:2020dpa} experiments, together with the associated $Q$ dependent effective exposure expressed in [g-day].}
	\label{tab:Qevents}
\end{table}

A second class of experiments of interest for this work includes EDELWEISS~\cite{EDELWEISS:2020fxc} and SENSEI~\cite{SENSEI:2020dpa}.~These experiments operate, respectively, germanium and silicon crystal detectors, and search for sub-GeV DM particles in events associated with the production of electron-hole pairs resulting from electronic transitions induced by DM-electron scattering in the detector crystals.~In order to compare our theoretical rate, Eq.~(\ref{eq:Robs}), with EDELWEISS~\cite{EDELWEISS:2020fxc} and SENSEI~\cite{SENSEI:2020dpa} results, we first calculate the total event rate per unit detector mass by dividing Eq.~(\ref{eq:Robs}) by the detector mass, $m_{\rm det}=N_{\rm cell} m_{\rm cell}$, where $m_{\rm cell}$ is the mass of a unit cell, 
\begin{align}
{\rm d }\widetilde{\mathscr{R}}_{\rm crystals} \equiv {\rm d }\mathscr{R}_{\rm theory}/m_{\rm det} \,,
\label{eq:Rcrystal_obs}
\end{align}
and then convert the deposited energy $\Delta E$ in Eq.~(\ref{eq:Rcrystal_obs}) into a number of electron-hole pairs produced in a DM-electron scattering event, which we denote by $Q$.~We convert $\Delta E$ into $Q$ using~\cite{Essig:2015cda}
\begin{align}
Q(\Delta E) = 1 + \lfloor (\Delta E - E_{\rm gap})/\varepsilon \rfloor\,,
\label{eq:conv}
\end{align}
where $\lfloor \cdot \rfloor$ is the floor function.~The observed band-gap and mean energy per electron-hole pair are $E_{\rm gap}=0.67$~eV and $\varepsilon=3.0$~eV for germanium~\cite{EDELWEISS:2020fxc}, and $E_{\rm gap}=1.20$~eV and $\varepsilon=3.8$~eV for silicon~\cite{SENSEI:2020dpa}.~We can then calculate the expected number of DM signal events corresponding to $Q=n$ electron-hole pairs, where $n\in \mathbb{N}$, by performing the $\Delta E$ integral in Eq.~(\ref{eq:Robs}) over the interval [$E_{\rm gap} +(n-1)\epsilon , E_{\rm gap}+n \epsilon$), and multiplying the result by the $Q$ dependent effective exposure (that is, the exposure times the detection efficiency).~We report the latter, together with the events actually observed by the EDELWEISS~\cite{EDELWEISS:2020fxc} and SENSEI~\cite{SENSEI:2020dpa} experiments in Tab.~\ref{tab:Qevents}.

Let us now denote by $\mathcal{S}_i$ the expected number of DM signal events in the $i$-th S2 bin, for XENON10 and XENON1T, or in the $i$-th $Q$ bin, for EDELWEISS and SENSEI.~Furthermore, let $\mathcal{D}_i$ be the number of actually observed events in the same (S2 or $Q$) bin.~For each of the four experiments considered here, we calculate 90\% confidence level (C.L.) exclusion limits on the strength of a given DM-electron interaction by requiring that $S_i\le S_i^{90\%}$ for all bins, where
\begin{align}
\sum_{k=\mathcal{D}_i+1}^{\infty} \mathcal{P}(k\,|\,\mathcal{S}_i^{90\%}) = 0.9 
\label{eq:stat}
\end{align}
defines $S^{90\%}_i$ and the number of counts in each bin is assumed to obey Poisson statistics.~Here, $ \mathcal{P}(k\,|\,\mathcal{S}_i^{90\%}) $ is the Poisson probability of observing $k$ events when an average of $\mathcal{S}_i^{90\%}$ is expected.

By computing exclusion limits using Eq.~(\ref{eq:stat}), we assume that the observed S2 or $Q$ events do not allow one to reject the ``null hypothesis'' in favour of a DM discovery.~For a detailed description of possible background contributions to $\mathcal{D}_i$, we refer to the experimental works~\cite{EDELWEISS:2020fxc,SENSEI:2020dpa,Angle:2011th,Aprile:2019xxb}.~For example, environmental backgrounds and dark current events from thermal excitations are the main contributions to $\mathcal{D}_i$ in the case of SENSEI.~The exclusion limits that we obtain from Eq.~(\ref{eq:stat}) are conservative, as they do not rely on any background subtraction.

\begin{table}[t]
	\centering
	\begin{tabular}{|ll|}
	\cline{1-2}
	\textbf{Simplified models and underlying Lorentz structures}			&	\\
	Lagrangian				&Lorentz structure	\\
	\cline{1-2}
	$\mathcal{L}_{g_4 h_4}^{(0)} = -ig_4(S^{\dagger}\partial_{\mu}S-\partial_{\mu}S^{\dagger}S)G^{\mu}-h_4(\bar{e}\gamma_{\mu}\gamma^5e)G^{\mu}$			&$\partial_{-}\otimes A$			\\
	$\mathcal{L}_{\lambda_3 h_4}^{(1/2)} = - \lambda_{3}\bar\chi\gamma^\mu\chi G_{\mu}-h_4\bar{e}\gamma_{\mu}\gamma^{5}eG^{\mu}$			& $V \otimes A$			\\
	$\mathcal{L}_{\lambda_4 h_3}^{(1/2)} = -\lambda_{4}\bar\chi\gamma^\mu\gamma^5\chi G_{\mu}-h_3\bar{e}\gamma_{\mu}eG^{\mu}$			&$A \otimes V$			\\
	$\mathcal{L}_{b_5 h_4}^{(1)} =-i b_{5} \left[X _{\nu}^{\dagger}\partial_{\mu}X^{\nu} - (\partial_{\mu}X^{\dagger\nu } )X_{\nu} \right]G^\mu - h_4 G_\mu\bar{e}\gamma^\mu\gamma^{5}e$		& $\partial_{-}\otimes A$				\\
	$\mathcal{L}_{\Re(b_6) h_3}^{(1)} = -\Re(b_{6}) \partial_\nu \left( X^{\dagger\nu } X_{\mu}+X^{\dagger}_{\mu} X^{\nu} \right)G^{\mu} -h_3G_\mu\bar{e}\gamma^\mu e$		&$\partial_{+}\otimes V$				\\
	$\mathcal{L}_{\Im(b_6) h_3}^{(1)} = -i\Im(b_{6}) \partial_\nu \left( X^{\dagger\nu } X_{\mu}-X^{\dagger}_{\mu} X^{\nu} \right)G^{\mu} -h_3G_\mu\bar{e}\gamma^\mu e$		&$\partial_{-}\otimes V$				\\
	$\mathcal{L}_{\Re(b_7) h_3}^{(1)} = -\Re(b_{7}) \epsilon_{\mu\nu\rho\sigma}  \left (X^{\dagger\mu}\partial^{\nu}X^{\rho} + X^{\mu}\partial^{\nu}X^{\dagger\rho }\right) G^{\sigma}
-h_3G_\mu\bar{e}\gamma^\mu e $		&$\epsilon\partial_{+}\otimes V$			\\
        $\mathcal{L}_{\Im(b_7) h_4}^{(1)} = -i \Im(b_{7}) \epsilon_{\mu\nu\rho\sigma}  \left (X^{\dagger\mu}\partial^{\nu}X^{\rho} - X^{\mu}\partial^{\nu}X^{\dagger\rho }\right) G^{\sigma}
 - h_4 G_\mu\bar{e}\gamma^\mu\gamma^{5}e$ &$\epsilon\partial_{-}\otimes A$ \\  
	\cline{1-2}
	\end{tabular}
	\caption{Lagrangians for a selection of simplified models with non-standard material response functions.}
	\label{tab:Lorentz}
\end{table}

\subsection{Constraints}
We are now ready to set constraints on the models of Sec.~\ref{sec:lagrangians} from the null result of XENON10, XENON1T, EDELWEISS and SENSEI.~While we have so far computed the DM response functions one would obtain when {\it all} coupling constants in a given Lagrangian from Sec.~\ref{sec:lagrangians} are different from zero (Eqs.~(\ref{eq:R00}), (\ref{eq:R01}), (\ref{eq:R1/20}), (\ref{eq:R1/21}), (\ref{eq:R10}) and (\ref{eq:R11})), here we restrict ourselves to a subset of ``simplified models'' each characterised by four parameters, namely:~the DM and mediator mass, a coupling constant for the DM-DM-mediator vertex, and a coupling constant for the electron-electron-mediator interaction vertex.~This choice will simplify the discussion while still allowing us to highlight the importance of the non-standard response functions, $\bar{\mathcal{W}}_l(q,\Delta E)$, $l\neq 1$, in the non-relativistic modelling of DM-electron scattering in materials.

Specifically, we focus on the simplified models defined in Tab.~\ref{tab:Lorentz}.~The corresponding Lagrangians follow by integration by parts from the Lagrangians in Sec.~\ref{sec:lagrangians} when only {\it one pair} of coupling constants at the time is assumed to be different from zero.~We focus on these eight simplified models as they generate at least one non-standard DM response function and, therefore, can only be investigated within the response function formalism we developed in~\cite{Catena:2019gfa,Catena:2021qsr}.

\begin{figure}[t]
\begin{center}
\begin{minipage}[t]{0.495\linewidth}
\centering
\includegraphics[width=\textwidth]{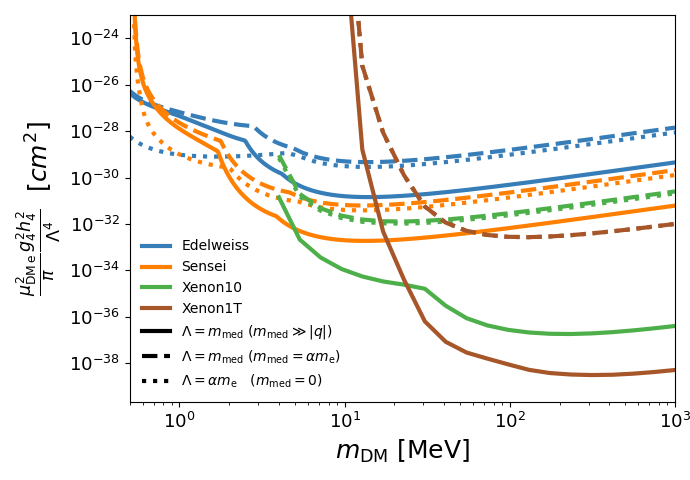}
\end{minipage}
\begin{minipage}[t]{0.495\linewidth}
\centering
\includegraphics[width=\textwidth]{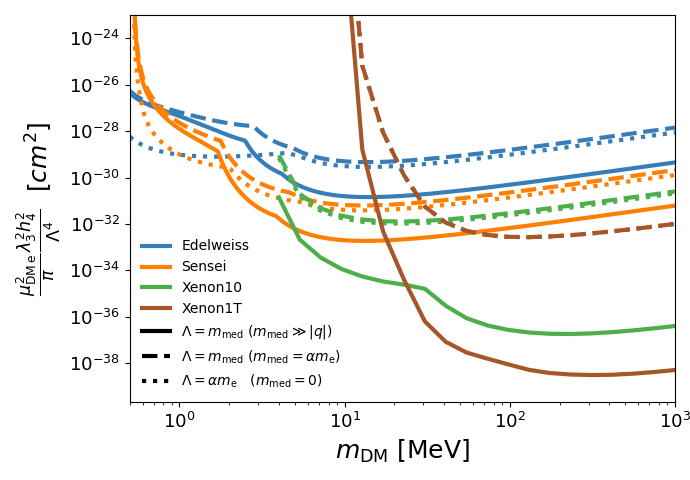}
\end{minipage}
\begin{minipage}[t]{0.495\linewidth}
\centering
\includegraphics[width=\textwidth]{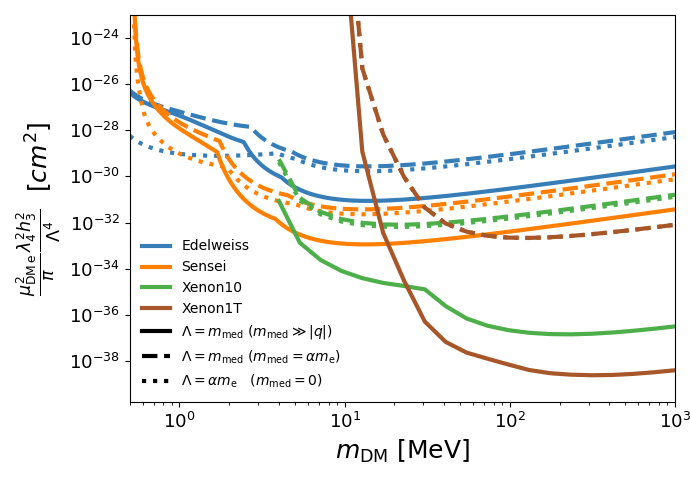}
\end{minipage}
\begin{minipage}[t]{0.495\linewidth}
\centering
\includegraphics[width=\textwidth]{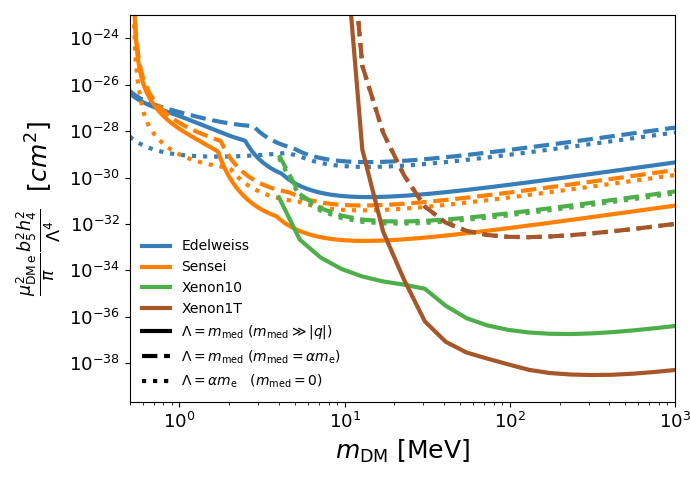}
\end{minipage}
\end{center}
\caption{90\% C.~L. exclusion limits on the reference cross section $\sigma_{\rm ref}$ for the simplified models with couplings constants $g_4$ and $h_4$ (top left panel), $\lambda_3$ and $h_4$ (top right panel), $\lambda_4$ and $h_3$ (bottom left panel) and $b_5$ and $h_4$ (bottom right panel).~See Tab.~\ref{tab:Lorentz} for the corresponding Lorentz structures.~Solid, dotted and dashed lines  refer to long-range, ``intermediate-range'' with $m_{\rm med}=\alpha m_e$ and long-range interactions, while blue, orange, green and maroon lines are associated with the EDELWEISS, SENSEI, XENON10 and XENON1T experiments, respectively.}
\label{fig:1}
\end{figure}

\begin{figure}[t]
\begin{center}
\begin{minipage}[t]{0.495\linewidth}
\centering
\includegraphics[width=\textwidth]{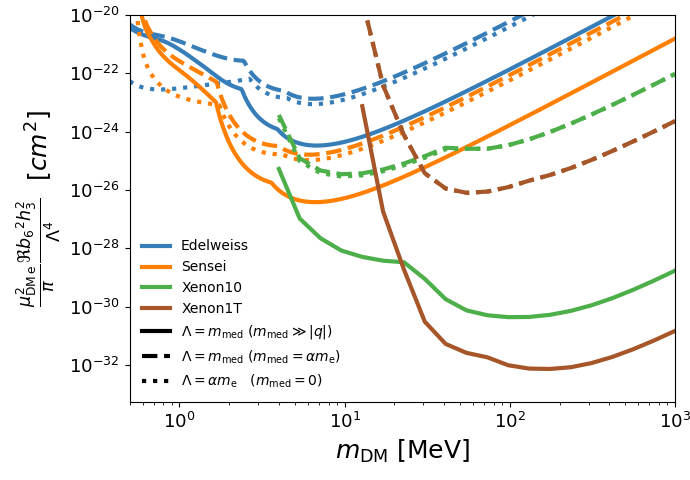}
\end{minipage}
\begin{minipage}[t]{0.495\linewidth}
\centering
\includegraphics[width=\textwidth]{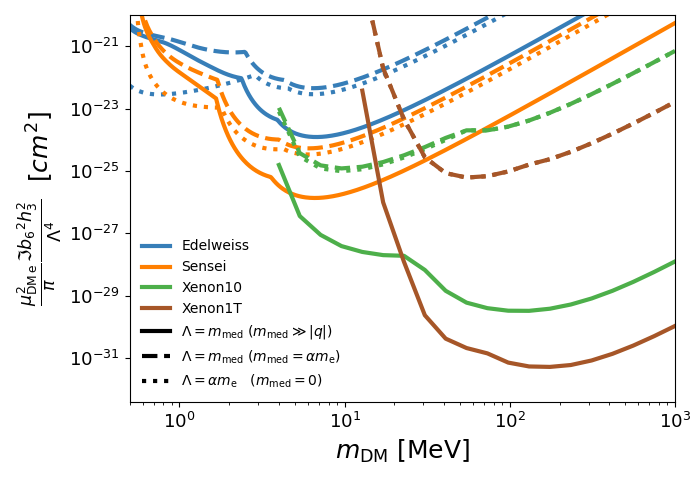}
\end{minipage}
\begin{minipage}[t]{0.495\linewidth}
\centering
\includegraphics[width=\textwidth]{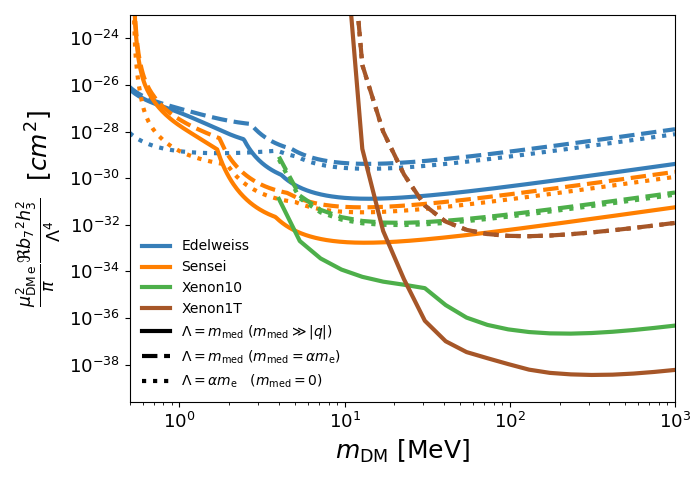}
\end{minipage}
\begin{minipage}[t]{0.495\linewidth}
\centering
\includegraphics[width=\textwidth]{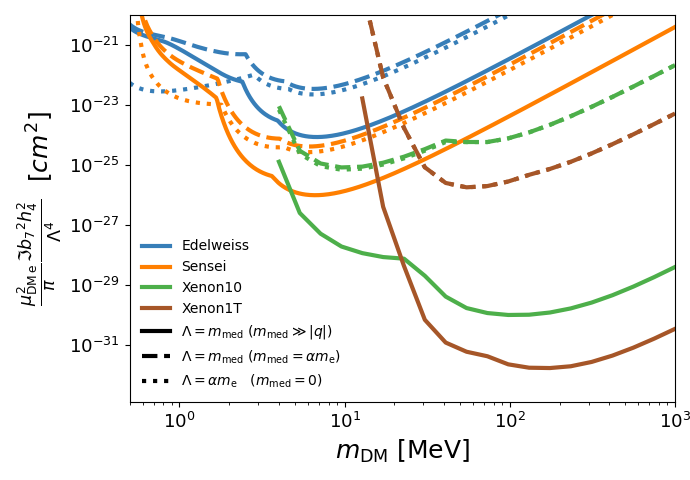}
\end{minipage}
\end{center}
\caption{Same as Fig.~\ref{fig:1}, for the simplified models with couplings constants $\Re(b_6)$ and $h_3$ (top left), $\Im(b_6)$ and $h_3$ (top right), $\Re(b_7)$ and $h_3$ (bottom left) and $\Im(b_7)$ and $h_4$ (bottom right).}
\label{fig:2}
\end{figure}

\begin{figure}[t]
\begin{center}
\begin{minipage}[t]{0.495\linewidth}
\centering
\includegraphics[width=\textwidth]{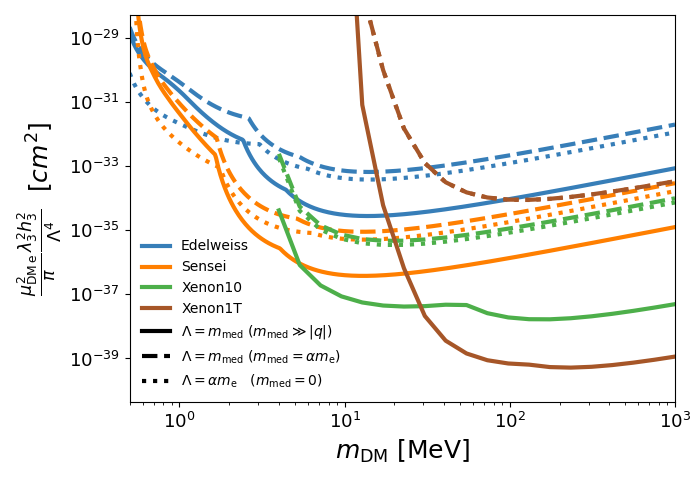}
\end{minipage}
\begin{minipage}[t]{0.495\linewidth}
\centering
\includegraphics[width=\textwidth]{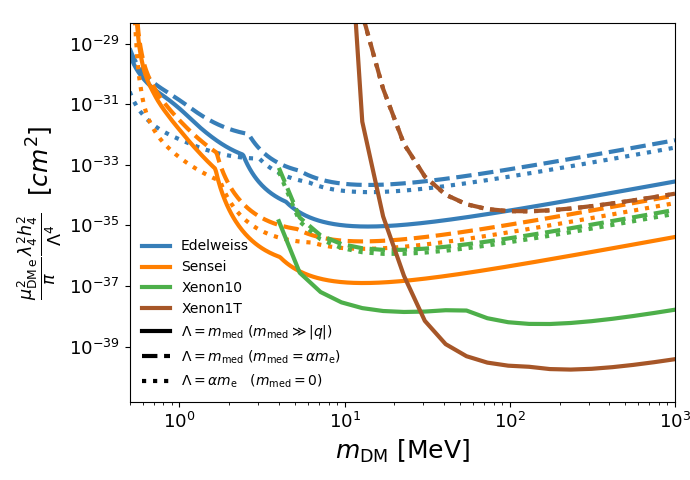}
\end{minipage}
\begin{minipage}[t]{0.495\linewidth}
\centering
\includegraphics[width=\textwidth]{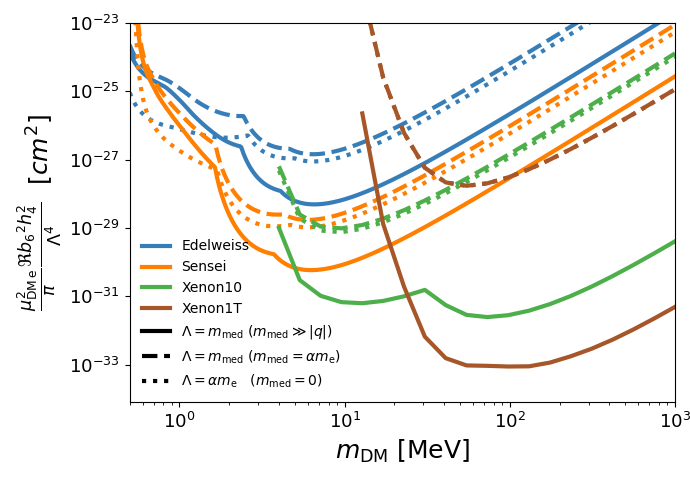}
\end{minipage}
\begin{minipage}[t]{0.495\linewidth}
\centering
\includegraphics[width=\textwidth]{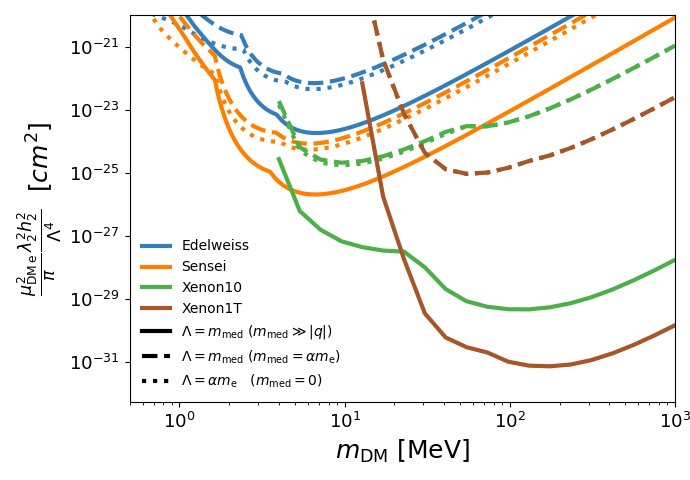}
\end{minipage}
\end{center}
\caption{Same as Fig.~\ref{fig:1}, for the simplified models defined by the couplings constants $\lambda_3$ and $h_3$ (top left), $\lambda_4$ and $h_4$  (top right), $\Re(b_6)$ and $h_4$ (bottom left) and $\lambda_2$ and $h_2$ (bottom right).}
\label{fig:3}
\end{figure}

The Lorentz structure of the Lagrangians in Tab.~\ref{tab:Lorentz} is variegated.~For example, $\mathcal{L}_{g_4 h_4}^{(0)}$ is characterised by a derivative coupling between scalar DM and a vector mediator, and an axial coupling between the vector mediator and the electron.~Symbolically, we denote this Lorentz structure by $\partial_{-} \otimes A$, where the minus sign in $\partial_{-} $ refers to the minus sign in parenthesis in the first term of $\mathcal{L}_{g_4 h_4}^{(0)}$.~Analogously, $\mathcal{L}_{\Re(b_7) h_3}^{(1)}$ is characterised by a derivative coupling between vector DM and a vector mediator, and a vector coupling between the mediator and the electron.~We denote this Lorentz structure by $\epsilon\partial_{+} \otimes V$ to emphasise that the derivative coupling in question includes a Levi-Civita tensor.~By adopting the notation outlined in these examples, Tab.~\ref{tab:Lorentz} shows the relation between our simplified models and the underlying Lorentz structures.

For the models in Tab.~\ref{tab:Lorentz}, we calculate 90\% C.~L. exclusion limits on the reference cross section,
\begin{align}
\sigma_{\rm ref} = \frac{\mu^2_{\rm DMe}}{\pi} \frac{\mathcal{P}^2_1 \mathcal{P}^2_2}{\Lambda^4}\,,
\label{eq:sigmaref}
\end{align}
from the null result reported by the XENON10~\cite{Angle:2011th}, XENON1T\cite{Aprile:2019xxb}, EDELWEISS~\cite{EDELWEISS:2020fxc} and SENSEI~\cite{SENSEI:2020dpa} collaborations.~In Eq.~(\ref{eq:sigmaref}), $\mu_{\rm DMe}$ is the DM-electron reduced mass, $\mathcal{P}_1$ and $\mathcal{P}_2$ are the coupling constants that define the given simplified model, while $\Lambda$ is a reference mass scale which depends on the assumed value for the mediator mass, $m_{\rm med}$.~For the simplified models in Tab.~\ref{tab:Lorentz}, we consider three cases for the mediator mass, each associated with a specific choice for $\Lambda$:~1) $m_{\rm med}\gg|\vec q|$, with $\Lambda=m_{\rm med}$, 2) $m_{\rm med}=\alpha m_e$, with $\Lambda=m_{\rm med}$, and, finally, 3) $m_{\rm med}=0$, with $\Lambda=\alpha m_e$.~The first (last) case corresponds to short-range (long-range) interactions, while the second one corresponds to $m_{\rm med}$ values that are comparable with a ``reference''  momentum transfer, which we set to $q_{\rm ref}=\alpha m_e$, where $\alpha$ is the fine structure constant.

Fig.~\ref{fig:1} shows the 90\% C.~L. exclusion limits on the reference cross section $\sigma_{\rm ref}$ that we obtain for the simplified models defined by the couplings constants $g_4$ and $h_4$ (top left panel), $\lambda_3$ and $h_4$ (top right panel), $\lambda_4$ and $h_3$ (bottom left panel) and $b_5$ and $h_4$ (bottom right panel).~In all panels, exclusion limits are presented as a function of the DM particle mass, as well as for different mediator masses and experiments.~More specifically, the solid, dotted and dashed lines in the figure refer to long-range, ``intermediate-range'' with $m_{\rm med}=\alpha m_e$ and long-range interactions, while blue, orange, green and maroon lines are associated with the EDELWEISS, SENSEI, XENON10 and XENON1T experiments, respectively, as indicated in the legends.

There are no appreciable differences in the exclusion limits presented in the four panels of Fig.~\ref{fig:1} because the underlying simplified models either generate the operator $\mathcal{O}_7$ only in the non-relativistic limit, or linear combinations of the interaction operators $\mathcal{O}_7$ and $\mathcal{O}_8$, and $\mathcal{O}_7$  and $\mathcal{O}_9$.~However, the operator $\mathcal{O}_7$ alone generates three non-standard material response functions.

As expected from Fig.~\ref{fig:Wratio1} and Fig.~\ref{fig:Wratio2}, and from the fact that xenon ionisation energies are larger than germanium and silicon band gaps by an order 10 factor, for $m_{\rm DM}$ below about 5~MeV crystal detectors such as EDELWEISS and SENSEI place stronger constraints on the cross section $\sigma_{\rm ref}$ than the xenon detectors XENON10 and XENON1T in all panels of Fig.~\ref{fig:1}.~In contrast, XENON10 and XENON1T set stronger bounds on $\sigma_{\rm ref}$ for values of the DM particle mass that are larger than about 5~MeV because of their larger experimental exposure. 

In the case of XENON10 and XENON1T, exclusion limits for short-range interactions are always stronger than for long-range interactions, except near mass threshold (i.e.~for $m_{\rm DM}$ close to the smallest testable DM mass), where they tend to overlap.~Furthermore, exclusion limits for short-range and intermediate-range interactions essentially coincide.~Both results are a consequence of the differential ionisation rate being proportional to $\sigma_{\rm ref}$, $\sigma_{\rm ref} (1+|\vec q|/q_{\rm ref})^{-4}$ and $\sigma_{\rm ref} q_{\rm ref}^4/|\vec q|^4$, for long-, intermediate- and short-range interactions, and $|\vec q|^4/q_{\rm ref}^4$ being typically larger than one for DM particle masses in the range probed by XENON10 and XENON1T.

In the case of EDELWEISS and SENSEI, the exclusion limits that we find in Fig.~\ref{fig:1} for $m_{\rm DM}$ larger than about 2 MeV and for different mediator masses exhibit the same hierarchy described above for XENON10 and XENON1T, the only difference being a non negligible separation between the exclusion limits obtained for short-range and intermediate-range interactions.~In contrast, for DM particle masses below about 2 MeV, long-range interactions are associated with significantly stronger exclusion limits than short- or intermediate-range interactions.~This result can be explained by recalling the aforementioned scaling with $|\vec q|/q_{\rm ref}$ of the differential ionisation or excitation rate, and by noticing that for crystal detectors $|\vec q|/q_{\rm ref}$ is smaller than one for sufficiently small DM particle masses, as one can infer from Figs.~\ref{fig:Wratio1} and \ref{fig:Wratio2}.

Fig.~\ref{fig:2} shows the 90\% C.~L. exclusion limits on the reference cross section $\sigma_{\rm ref}$ that we find for the simplified models with couplings constants $\Re(b_6)$ and $h_3$ (top left panel), $\Im(b_6)$ and $h_3$ (top right panel), $\Re(b_7)$ and $h_3$ (bottom left panel) and $\Im(b_7)$ and $h_4$ (bottom right panel).~The exclusion limits in Fig.~\ref{fig:2} exhibit the same hierarchies and patterns discussed in details for Fig.~\ref{fig:1}.~Importantly, none of the simplified models underlying Figs.~\ref{fig:1} and \ref{fig:2} could have been investigated accurately without having first computed the material response functions $\bar{\mathcal{W}}_2(q,\Delta E)$ and $ \bar{\mathcal{W}}_3(q,\Delta E)$ for atoms and crystals.~Furthermore, we find that for vector DM models with a vector mediator also the material response function $\bar{\mathcal{W}}_4(q,\Delta E)$ is crucial.

We conclude this analysis by validating our results through a comparison with the exclusion limits associated with models that generate the material response function $\bar{\mathcal{W}}_1(q,\Delta E)$ only.~Fig.~\ref{fig:3} shows our 90\% C.~L. exclusion limits on the reference cross section $\sigma_{\rm ref}$ for the simplified models that are identified by the couplings constants $\lambda_3$ and $h_3$ (top left panel), $\lambda_4$ and $h_4$  (top right panel), $\Re(b_6)$ and $h_4$ (bottom left panel) and $\lambda_2$ and $h_2$ (bottom right panel).~The top panels in this figure correspond to the familiar spin-independent and spin-dependent interactions, arising from the $\mathcal{O}_1$ and $\mathcal{O}_4$ operators, respectively.~The associated exclusion limits presented here agree with previous results in the literature, see e.g.~\cite{Catena:2019gfa,Catena:2021qsr}.~Notice that the ($\lambda_3, h_3$) case coincides with a dark photon model with Dirac DM.~The bottom panels correspond to models that generate the $\bar{\mathcal{W}}_1(q,\Delta E)$ material response function only, but with a DM response function scaling with $|\vec q|^2$ (left panel) and $|\vec q|^4$ (right panel), respectively.

\section{Summary and conclusion}
\label{sec:conclusion}

A still open question in the field of sub-GeV DM direct detection is whether experimental results can be interpreted within a theoretical framework where the response of detector materials is described in terms of a single ionisation or crystal form factors only, as in the case of the dark photon model.~In this work, we have addressed this question by computing the rate of electronic transitions induced by DM-electron scattering in isolated xenon atoms, as well as in silicon and germanium crystals, for a variety of models where the DM particle can have spin 0, spin 1/2 or spin 1, while the particle that mediates the DM-electron interaction can have either spin 0 or spin 1.~We have found several examples for which an accurate description of the non-relativistic scattering of DM particles by the electrons bound in detector materials requires material response functions that go beyond the standard ionisation and crystal form factors.~For simplicity, we have illustrated this conclusion by restricting ourselves to the case of ``simplified model'', where just two coupling constants at the time are different from zero.~However, the fact that non-standard material response functions can arise in ``minimal extensions'' of the dark photon model is a general conclusion of our study, which does not rely on this restriction.~This result corroborates our previous findings~\cite{Catena:2019gfa,Catena:2021qsr}, and shows the importance of a response function formalism in the interpretation of future DM direct detection data.  

For the eight models that generate a non-standard material response, we have calculated 90\% C.~L. exclusion limits on the reference cross section for DM-electron electron scattering, Eq.~(\ref{eq:sigmaref}), from the null result reported by the   XENON10~\cite{Angle:2011th}, XENON1T\cite{Aprile:2019xxb}, EDELWEISS~\cite{EDELWEISS:2020fxc} and SENSEI~\cite{SENSEI:2020dpa} experiments.~We have performed this calculation as a function of the DM particle mass and for different values of the mediator mass.~We have not only considered small and large mediator masses, corresponding to contact and long range interactions, but also scenarios in which the mediator mass is comparable with the momentum transfer in the scattering.~Importantly, none of these exclusion limits could have accurately been extracted from data without having first computed the non-standard material response functions of Sec.~\ref{sec:dd}.

A byproduct of our analysis is Eq.~(\ref{eq:Wcompact}) which, expanding on our previous work~\cite{Catena:2019gfa,Catena:2021qsr}, has enabled us to express the response of xenon and crystal detectors to an external DM probe in the same, compact form (which was not the case in~\cite{Catena:2019gfa,Catena:2021qsr}).~An interesting aspect of this equation is that it allows for a simple comparison of the performance of different materials in the context of sub-GeV DM direct detection.~We have applied this result to compare the response of xenon, silicon and germanium detectors finding that these materials perform best in complementary regions of the plane spanned by the momentum transfer and the deposited energy.
 
 To conclude, we give a positive answer to the question of whether new material response functions beyond the standard crystal and ionisation form factors are needed for the interpretation of present and future direct detection experiments searching for DM-induced electronic transitions.

\acknowledgments 
RC and TE acknowledge that this work was produced as part of the Knut and Alice Wallenberg project Light Dark Matter (Dnr. KAW 2019.0080).~RC also acknowledges support from an individual research grant from the Swedish Research Council (Dnr.~2018-05029).~TE thanks the Theoretical Subatomic Physics group at Chalmers University of Technology for its hospitality.~NAS and MM were supported by the ETH Zurich, and by the European Research Council (ERC) under the European Union's Horizon 2020 research and innovation programme project HERO Grant Agreement No. 810451.

\appendix
\section{Squared transition amplitudes}
\label{app:details}
Here, we provide the details underling the derivation of Eqs.~(\ref{eq:R1/20}), (\ref{eq:R1/21}), (\ref{eq:R10}) and (\ref{eq:R11}).~These four equations correspond to the DM response functions for, respectively, fermionic dark matter with a scalar mediator, fermionic dark matter with a vector mediator, vector dark matter with a scalar mediator and, finally, vector dark matter with a vector mediator.

\subsection{Fermionic dark matter with a scalar mediator}
The free amplitude for DM-electron scattering in the case of spin-1/2 DM with a scalar mediator is
\begin{align}
    i\mathcal{M} 
    &=   
            \Bar{u}^{s'}_\chi(p') i(\lambda_1 + i\lambda_2\gamma^5)u_\chi^s(p)
             \frac{i}{(p-p')^2-m_\phi^2}
             \Bar{u}_e^{r'}(k')i(h_1 + ih_2\gamma^5)u_e^r(k)\,,
        \label{eq:M1/20}
\end{align}
where $u_e^r$ and $u^s_\chi$ are electron and DM free spinors, while $k$ and $k'$ ($p$ and $p'$) are the initial and final electron (DM) four-momenta.~By using the spinor bilinear expansions in App.~\ref{app:identities}, we find that in the non-relativistic limit the amplitude $i\mathcal{M}$ in Eq.~(\ref{eq:M1/20}) can be written as follows
\begin{align}
i\mathcal{M} &\simeq
i4m_\chi m_e
    \left(
    \frac{\lambda_1h_1}{|\vec q\,|^2+m_\phi^2} \langle \mathcal{O}_{1} \rangle +
    \frac{\lambda_2h_2}{|\vec q\,|^2+m_\phi^2}\frac{m_e}{m_\chi} \langle \mathcal{O}_6 \rangle  
    -\frac{\lambda_1h_2}{|\vec q\,|^2+m_\phi^2} \langle \mathcal{O}_{10} \rangle \right. \nonumber\\
 &+ \left.  \frac{\lambda_2h_1}{|\vec q\,|^2+m_\phi^2}\frac{m_e}{m_\chi} \langle \mathcal{O}_{11} \rangle
    \right) \,.
\label{eq:M1/20nr}
\end{align}
The operators appearing in angle brackets in this expression are listed in Tab.~\ref{tab:operators}.~From Eq.~(\ref{eq:M1/20nr}), we find

\begin{align}
\overline{| \mathcal{M}|^2} | f_{1\rightarrow 2}|^2  &= \left( c_1^2 + \frac{1}{16}c_6^2\frac{\qAbs^4}{m_e^4} + \frac{1}{4}c_{10}^2\frac{\qAbs^2}{m_e^2} + \frac{1}{4}c_{11}^2\frac{\qAbs^2}{m_e^2}\right)  | f_{1\rightarrow 2}|^2\nonumber\\ 
2 m_e \Re \left\{ \overline{\mathcal{M}f_{1\rightarrow 2} [\nabla_{\boldsymbol{\ell}} \mathcal{M}^{*}] \cdot \boldsymbol{f}^{*}_{1\rightarrow 2} } \right\} 
&= 0 \nonumber\\
 m_e^2\overline{|\nabla_{\boldsymbol{\ell}} \mathcal{M}(\vec q,\vec v^\perp_{\rm el}) ] \cdot \boldsymbol{f}_{1\rightarrow 2}(\vec q) |^2} &= 0 \,,
\label{eq:M2S1/20}
\end{align}
which implies Eq.~(\ref{eq:R1/20}), once the azimuthal average in Eq.~(\ref{eq:azimuthal}) is computed by using Eq.~(\ref{eq:azimuthal_examples}).~We give the relation between the non-relativistic coupling constants in the equations above, i.e. $c_1$, $c_6$, $c_{10}$ and $c_{11}$, and the  particle masses and coupling constants in Eq.~(\ref{eq:M1/20nr}) in Tab.~\ref{tab:S0.5coeff}.~Here and below, we implicitly assume that $\mathcal{M}$ bilinears are evaluated at $\boldsymbol{\ell}=0$ and $\cos\theta = \xi$.

\subsection{Fermionic dark matter with a vector mediator}
In the case of spin-1/2 DM with a vector mediator, the free amplitude for DM-electron scattering is
\begin{align}
i\mathcal{M}
  &=
   \left[ 
        \Bar{u}^{s'}(p')_\chi i(\lambda_3\gamma_\mu + \lambda_4\gamma_\mu\gamma^5)u_\chi^s(p)
    \right]
        \frac{-ig^{\mu\nu}}{(p-p')^2-m_G^2}
    \left[ 
        \Bar{u}_e^{r'}(k')i(h_3\gamma_\nu + h_4\gamma_\nu\gamma^5)u_e^r(k)
    \right] \,,
\end{align}
which, reduces to
\begin{align}
i\mathcal{M}
  &\simeq
  i4m_\chi m_e
    \Bigg[
        -\frac{\lambda_3h_3}{|\vec q|^2 + m_G^2}\NRop{1} +
        \frac{4\lambda_4 h_4}{|\vec q|^2 + m_G^2}\NRop{4} + 
        \frac{2\lambda_3h_4}{|\vec q|^2 + m_G^2}\NRop{7}
        -\frac{2\lambda_4h_3}{|\vec q|^2 + m_G^2}\NRop{8}\nonumber\\
        & + 
        \left(
        \frac{2\lambda_3h_4}{|\vec q|^2 + m_G^2}\frac{m_e}{m_\chi} + 
        \frac{2\lambda_4h_3}{|\vec q|^2 + m_G^2}
        \right)\NRop{9}
        \label{eq:M1/21nr}
    \Bigg] \,,
\end{align}
in the non-relativistic limit.~We give explicit expressions for the operators and coupling constants in this equation in Tab.~\ref{tab:operators} and Tab.~\ref{tab:S0.5coeff}, respectively.~Finally, by applying Eq.~(\ref{eq:M1/21nr}), we obtain
\begin{align}
\overline{| \mathcal{M}|^2} | f_{1\rightarrow 2}|^2  &= \left( c_1^2 + \frac{3}{16}c_4^2 + \frac{1}{4}c_7^2\vAbs^2 + \frac{1}{4} c_8^2 |\vec v^\perp_{\rm el}|^2 + \frac{1}{8}c_9^2\frac{\qAbs^2}{m_e^2} \right)  | f_{1\rightarrow 2}|^2\nonumber\\ 
2 m_e \Re \left\{ \overline{\mathcal{M}f_{1\rightarrow 2} [\nabla_{\boldsymbol{\ell}} \mathcal{M}^{*}] \cdot \boldsymbol{f}^{*}_{1\rightarrow 2} } \right\} 
&= -\left(\frac{1}{2}c_7^2 + \frac{1}{2}c_8^2\right)\vvp\cdot\vA \nonumber\\
 m_e^2\overline{|\nabla_{\boldsymbol{\ell}} \mathcal{M}(\vec q,\vec v^\perp_{\rm el}) ] \cdot \boldsymbol{f}_{1\rightarrow 2}(\vec q) |^2} &=
 \left(\frac{1}{4}c_7^2 + \frac{1}{4}c_8^2\right) |\cvff|^2\,.
\end{align}
By taking the azimuthal average of the equations above, we obtain the DM response functions in Eq.~(\ref{eq:R1/21}).

\subsection{Vector dark matter with a scalar mediator}
The free amplitude for DM-electron scattering in the case of spin-1 DM with a scalar mediator is
\begin{align}
 i\mathcal{M}
    &= 
    \left[ 
         i(b_1m_Xg^{\mu\nu})  \epsilon^{*s'}_\mu(p')\epsilon^s_\nu(p)
    \right]
        \frac{i}{(p-p')^2 - m_\phi^2}
    \left[ 
        \Bar{u}_e^{r'}(k')i(h_1 + ih_2\gamma^5)u_e^r(k)
    \right]\,,
\end{align}
and, in the non-relativistic limit,
\begin{align}
 i\mathcal{M}
    &\simeq
    i4m_Xm_e
    \left( 
    \frac{1}{2}\frac{b_1h_1}{|\mathbf{q}|^2+m_\phi^2}\NRop{1} - 
    \frac{1}{2}\frac{b_1h_2}{|\mathbf{q}|^2+m_\phi^2}\NRop{10} 
    \right) \,,
    \label{eq:M10nr}
\end{align}
from which
\begin{align}
\overline{| \mathcal{M}|^2} | f_{1\rightarrow 2}|^2  &= \left( c_1^2 + \frac{1}{4}c_{10}^2\frac{\qAbs^2}{m_e^2} \right)  | f_{1\rightarrow 2}|^2\nonumber\\ 
2 m_e \Re \left\{ \overline{\mathcal{M}f_{1\rightarrow 2} [\nabla_{\boldsymbol{\ell}} \mathcal{M}^{*}] \cdot \boldsymbol{f}^{*}_{1\rightarrow 2} } \right\} 
&= 0 \nonumber\\
 m_e^2\overline{|\nabla_{\boldsymbol{\ell}} \mathcal{M}(\vec q,\vec v^\perp_{\rm el}) ] \cdot \boldsymbol{f}_{1\rightarrow 2}(\vec q) |^2} &= 0\,.
 \label{eq:MA10}
\end{align}
The operators $\mathcal{O}_1$ and $\mathcal{O}_{10}$ are defined in Tab.~\ref{tab:operators}, whereas an explicit relation between $c_1$ and $c_{10}$ and the masses and coupling constants in Eq.~(\ref{eq:M10nr}) can be found in Tab.~\ref{tab:S1coeff}.~As a result, Eq.~(\ref{eq:R10}) follows from Eq.~(\ref{eq:MA10}) by a straightforward application of the definition~(\ref{eq:azimuthal}) and Eq.~(\ref{eq:azimuthal_examples}).

\subsection{Vector dark matter with a vector mediator}
Finally, in the case of spin-1 DM with a vector mediator the free amplitude for DM-electron scattering is
\begin{align}
i\mathcal{M}=&\bigg[-i b_5 \left(p^{\prime\mu}+ p^\mu\right) \epsilon^{s'\nu*}(p\,')\epsilon^{s}_\nu(p) \nonumber\\
&+\Re(b_6)\left(p_\nu' - p_\nu\right)\left(\epsilon^{s'\nu*}(p\,')\epsilon^{s\mu}(p)+\epsilon^{s'\mu*}(p\,')\epsilon^{s\nu}(p)\right) \nonumber\\
&+i\Im(b_6)\left(p_\nu' - p_\nu\right)\left(\epsilon^{s'\nu*}(p\,')\epsilon^{s\mu}(p)-\epsilon^{s'\mu*}(p\,')\epsilon^{s\nu}(p)\right) \nonumber\\
&-\Re(b_7)\varepsilon_{\mu\nu\rho\sigma}\left(p^{\prime\nu}+ p^\nu\right)\epsilon^{s'\mu*}(p\,')\epsilon^{s\rho}(p) +i\Im(b_7)\varepsilon_{\mu\nu\rho\sigma}\left(p^{\prime\nu}- p^\nu\right)\epsilon^{s'\mu*}(p\,')\epsilon^{s\rho}(p) \bigg] \nonumber\\
&\times \left[ \frac{-1}{(p-p')^2-m_G^2}\right] \ubar_e^{r'}(k')\gamma_\mu (h_3+h_4\gamma_5)u_e^r(k) \,.
\label{eq:M11}
\end{align}
By using the results listed in App.~\ref{app:identities}, we find that in the non-relativistic limit Eq.~(\ref{eq:M11}) reduces to
\begin{align}
i\mathcal{M} 
    \simeq   i 4m_Xm_e
    &\Bigg[
    -\frac{b_5h_3}{|\vec q\,|^2+m_G^2}
                    \NRop{1} +
    \left(
    \frac{\Im(b_6)h_3}{|\vec q\,|^2+m_G^2}\frac{m_e}{2m_X}\frac{|\vec q\,|^2}{m_e^2} 
   -
    \frac{2\Re(b_7)h_4}{|\vec q\,|^2+m_G^2}
    \right)
                    \NRop{4} 
    \nonumber\\
    &  + 
    \frac{\Im(b_6)h_3}{|\vec q\,|^2+m_G^2}\frac{m_e}{2m_X}
                    \NRop{5} -
    \frac{\Im(b_6)h_3}{|\vec q\,|^2+m_G^2}\frac{m_e}{2m_X}
                    \NRop{6} + 
    \frac{2b_5h_4}{|\vec q\,|^2+m_G^2}
                    \NRop{7} 
    \nonumber\\
    & +
    \frac{\Re(b_7)h_3}{|\vec q\,|^2+m_G^2}
                    \NRop{8} 
    +
    \left(               
    \frac{\Im(b_6)h_4}{|\vec q\,|^2+m_G^2}\frac{m_e}{m_X}
    -
    \frac{\Re(b_7)h_3}{|\vec q\,|^2+m_G^2}
    \right)
                    \NRop{9} 
    \nonumber\\
    & +
    \frac{1}{2}\frac{\Im(b_7)h_3}{|\vec q\,|^2+m_G^2}\frac{m_e}{m_X}
                    \NRop{11} -
    \frac{\Im(b_7)h_4}{|\vec q\,|^2+m_G^2}\frac{m_e}{m_X}
                    \NRop{14} 
    \nonumber\\
    & -
    \frac{\Re(b_6)h_3}{|\vec q\,|^2+m_G^2}\frac{m_e}{m_X} 
                    \NRop{17} +
    \frac{2\Re(b_6)h_4}{|\vec q\,|^2+m_G^2}\frac{m_e}{m_X} 
                    \NRop{18} +
    \frac{\Im(b_6)h_3}{|\vec q\,|^2+m_G^2}\frac{m_e^2}{2m_X^2}
                    \NRop{19} 
     \nonumber\\
    &- 
    \left(
    \frac{\Re(b_6)h_3}{|\vec q\,|^2+m_G^2}\frac{m_e}{m_X} 
    +
    \frac{\Im(b_7)h_4}{|\vec q\,|^2+m_G^2}\frac{m_e^2}{m_X^2}
    \right)
                    \NRop{20}
    \Bigg] \,.
    \label{eq:M11nr}
\end{align}
As above, we define the operators in Eq.~(\ref{eq:M11nr}) in Tab.~\ref{tab:operators}.~From Eq.~(\ref{eq:M11nr}), we obtain the expressions 
\begin{align}
\overline{| \mathcal{M}|^2} | f_{1\rightarrow 2}|^2  &=  \Bigg(
     c_1^2 + 
    \frac{1}{2}c_4^2 + 
    \frac{2}{3}c_5^2\big|\frqm\times\vvp\big|^2 + 
    \frac{1}{6}c_6^2\frac{\qAbs^4}{m_e^4} +
    \frac{1}{4}c_7^2\vAbs^2  \nonumber\\
    & +
    \frac{2}{3}c_8^2\vAbs^2 +\frac{1}{3}c_9^2\frac{\qAbs^2}{m_e^2} + 
    \frac{2}{3}c_{11}^2\frac{\qAbs^2}{m_e^2} + 
    \frac{1}{6}c_{14}^2\vAbs^2\frac{\qAbs^2}{m_e^2} \nonumber\\
    & + 
    \frac{1}{6}c_{17}^2\Big(\big|\frqm\cdot\vvp\big|^2 + \frac{\qAbs^2}{m_e^2}\vAbs^2\Big) + 
    \frac{1}{6}c_{18}^2\frac{\qAbs^2}{m_e^2}  \nonumber \\
    & +
    \frac{1}{3}c_{19}^2\frac{\qAbs^4}{m_e^4} +
    \frac{1}{12}c_{20}^2\frac{\qAbs^4}{m_e^4} + 
    \frac{1}{3}c_4c_6\frac{\qAbs^2}{m_e^2} +  \frac{2}{3}c_1c_{19}\frac{\qAbs^2}{m_e^2}\Bigg)
  | f_{1\rightarrow 2}|^2\nonumber\\ 
2 m_e \Re \left\{ \overline{\mathcal{M}f_{1\rightarrow 2} [\nabla_{\boldsymbol{\ell}} \mathcal{M}^{*}] \cdot \boldsymbol{f}^{*}_{1\rightarrow 2} } \right\} 
&=\left[ \frqm\cdot\vvp \left(\frac{4}{3}c_5^2 - \frac{1}{3}c_{17}^2 \right) \right] \frqm\cdot\vA \nonumber\\
    & - 
   \left[\frac{1}{2}c_7^2 + \frac{4}{3}c_8^2 + \frac{\qAbs^2}{m_e^2}\left(\frac{4}{3}c_5^2 + \frac{1}{3}c_{14}^2 + \frac{1}{3}c_{17}^2\right) \right]  \nonumber\\ &\times  \vvp\cdot\vA
\nonumber\\
 m_e^2\overline{|\nabla_{\boldsymbol{\ell}} \mathcal{M}(\vec q,\vec v^\perp_{\rm el}) ] \cdot \boldsymbol{f}_{1\rightarrow 2}(\vec q) |^2} &=
 \left[ \frac{1}{4}c_7^2 + \frac{2}{3}c_8^2 + \frac{\qAbs^2}{m_e^2}\left(\frac{2}{3} c_5^2 + \frac{1}{6}c_{14}^2 + \frac{1}{6}c_{17}^2\right)\right] \fAbs^2\nonumber\\
    & + 
   \left(-\frac{2}{3} c_5^2 + \frac{1}{6}c_{17}^2\right)  \left|\frqm\cdot \cvff\right|^2\,,
   \label{eq:MA11}
\end{align}
from which Eq.~(\ref{eq:R11}) follows by computing an azimuthal average, as shown explicitly in Eq.~(\ref{eq:azimuthal_examples}).~As in the previous examples, the coupling constants in Eq.~(\ref{eq:MA11}) are defined in Tab.~\ref{tab:S1coeff}.

\section{Useful identities}
\label{app:identities}
Below, we list expressions that we use in the non-relativistic reduction of free scattering amplitudes, and in the calculation of the squared modulus of the transition amplitude for the models in Sec.~\ref{sec:lagrangians}.~We start from the non-relativistic reduction of free spinors and spinor bilinears,
\begin{align}
    u_\chi^r(p) \simeq \frac{1}{\sqrt{4m_\chi}}
    \begin{pmatrix}
    (2m_\chi - \vec p \cdot \vec \sigma) \xi^r\\
    (2m_\chi + \vec p \cdot \vec \sigma) \xi^r\\
    \end{pmatrix} \,, \nonumber
\end{align}
and
\begin{align}
    \Bar{u}_\chi^{s'}(p') u_\chi^s(p)                             &\simeq 2m_\chi \, \xi^{\dagger s'}\xi^s \\
    \Bar{u}_\chi^{s'}(p')i\gamma^5 u_\chi^s(p)                    &\simeq 2i\vec q\cdot \vec s^{s's} \\
    \Bar{u}_\chi^{s'}(p')\gamma^\mu u_\chi^s(p)                   &\simeq \begin{pmatrix}
                                                                2m_\chi \, \xi^{\dagger s'}\xi^s \\
                                                                \vec P \, \xi^{\dagger s'}\xi^s + 2i\vec q\times \vec s^{s's}
                                                                \end{pmatrix} \\
    \Bar{u}_\chi^{s'}(p')\gamma^\mu\gamma^5 u_\chi^s(p)           &\simeq \begin{pmatrix}
                                                                2 \vec P\cdot \vec s^{s's} \\
                                                                4m_\chi \, \vec s^{s's}
                                                                \end{pmatrix} \\
    \Bar{u}_\chi^{s'}(p')\sigma^{\mu\nu} u_\chi^s(p)              &\simeq \begin{pmatrix}
                                                                0 & i\vec q\,\xi^{\dagger s'}\xi^s - 2 \vec P \times \vec s^{s's}\\
                                                                -i\vec q\,\xi^{\dagger s'}\xi^s + 2 \vec P \times \vec s^{s's} & 4m_\chi\,\epsilon_{ijk}\,s^{s'sk}
                                                                \end{pmatrix} \,,
\end{align}
respectively, which apply to the case of fermionic DM.~Here,  $\vec s^{s's} \equiv \xi^{\dagger s'}\vec{S}_{\rm DM}\xi^s$, $\vec P = \vec p + \vec p'$, $\vec q = \vec p - \vec p'$, $\xi^s$, $s=1,2$ and  $\xi^{\dagger s'}$, $s'=1,2$ are two component spinors, while $\vec \sigma = (\sigma_1, \sigma_2, \sigma_3)$ is a vector the component of which are the three Pauli matrices.~Similar expressions apply to the free electron spinor $u_e^r$.~In the non-relativistic reduction of scattering amplitudes, we also use the polarisation vectors of massive vector DM or spin-1 mediator particles.~They can be expanded as
\begin{align}
\epsilon^{s\mu}_\mu(\vec p)  \simeq \begin{pmatrix} \frac{1}{2m_{X}}(\vec P + \vec q\,)\cdot \vec e_s \\ \vec e_s \end{pmatrix}
\quad \quad 
\epsilon^{s'\mu*}(\vec p')  \simeq \begin{pmatrix} \frac{1}{2 m_X}(\vec P - \vec q\,)\cdot \vec e'_{s'} \\ \vec e'_{s'} \end{pmatrix}\,,
\end{align}
where $s,s'=1,2,3$, $m_X$ can be either the DM or the mediator mass, $\vec e_3=\vec p/|\vec p|$, $\vec e'_{3'}=\vec p' /|\vec p'|$, and $\vec e_s \cdot \vec e'_{s'}\simeq \delta_{ss'}$.~In addition, we use the non-relativistic expansion of the momentum transfer four-vector,
\begin{align}
q\equiv(E_{\vec p}-E_{\vec p'},\vec p - \vec p') \simeq \left(\frac{\vec P \cdot \vec q}{2m_{\rm DM}}  \, , \, \vec q\right) \,,
\end{align}
which allows us to write the non-relativistic reduction of scattering amplitudes in terms of $\vec q$ and 
\begin{align}
    \vvp &= \frac{\vec p+\vec p\,'}{2m_{\rm DM}} - \frac{\vec k+\vec k\,'}{2m_e} = \vec v - \frac{\vec k}{m_e} - \frac{\vec q}{2\mu_{\rm DM e}}\,,
\end{align}
where $\vec k$ ($\vec k'$) is the initial (final) free electron momentum, and $\vec v$ the DM-electron relative velocity.~We conclude this appendix by listing useful spin sums for electrons or spin-1/2 DM particles,
\begin{align}
    &\frac{1}{2}\sum_{r'r} 
        |\vec s^{r'r}|^2 
            = \frac{3}{4} 
    &\frac{1}{2}\sum_{r'r} (\vec s^{r'r}\times \vec A)\cdot \vec s^{r'r} = 0 \\
    &\frac{1}{2}\sum_{r'r} 
        (\vec s^{r'r}\cdot\vec A)
        (\vec s^{r'r}\cdot\vec B)  
            = \frac{1}{4}\vec A\cdot \vec B 
    &\frac{1}{2}\sum_{r'r} (\vec s^{r'r}\cdot\vec A)\delta^{r'r} = 0 \\
    &\frac{1}{2}\sum_{r'r} 
        (\vec s^{r'r}\times\vec A)
            (\vec s^{r'r}\times \vec B) 
         = \frac{1}{2}\vec A\cdot\vec B
    &\frac{1}{2}\sum_{r'r} (\vec s^{r'r}\times\vec A)\delta^{r'r} = 0 
\end{align}
and for spin-1 DM particles,
\begin{align}
    &\frac{1}{3}\sum_{s's} \vec A \cdot \vec s_V^{s's} = 0 \\
    &\frac{1}{3}\sum_{s's} \vec A \cdot \mathcal{S}^{s's} \cdot \vec s_V^{s's} = 0 \\
    &\frac{1}{3}\sum_{s's} (\vec A \cdot \vec s_V^{s's} )(\vec B \cdot \vec s_V^{s's} ) = \frac{2}{3}\vec A\cdot\vec B \\
    &\frac{1}{3}\sum_{s's} (\vec A \times \vec s_V^{s's} )\cdot(\vec B \times \vec s_V^{s's} ) = \frac{4}{3}\vec A\cdot\vec B \\
    &\frac{1}{3}\sum_{s's} (\vec A \cdot \mathcal{S}^{s's} \cdot \vec B)\delta^{s's} = \frac{1}{3}\vec A\cdot\vec B \\
    &\frac{1}{3}\sum_{s's} (\vec A \cdot \mathcal{S}^{s's} \cdot \vec B)  (\vec s_V^{s's}\cdot \vec C) = 0  \\
    &\frac{1}{3}\sum_{s's} (\vec A \cdot \mathcal{S}^{s's} ) \cdot (\vec B \cdot \mathcal{S}^{s's}) = \frac{2}{3}\vec A \cdot\vec B \\
    &\frac{1}{3}\sum_{s's} 
        |\vec A \cdot \mathcal{S}^{s's} \cdot \vec B|^2 
            = \frac{1}{6}|\vec A\cdot\vec B|^2 
                + \frac{1}{6}|\vec A|^2|\vec B|^2 \,,
\end{align}
where $\vec A$ and $\vec B$ are arbitrary three-dimensional vectors, $r=1,2$, $r'=1,2$, $s=1,2,3$, $s'=1,2,3$, and
\begin{align}
\vec s_V^{\ s's} &\equiv -i \vec e'_{s'}\times \vec e_{s} \\
\mathcal{S}^{s's}_{ij}&\equiv\frac{1}{2}(e_{si} e'_{s'j} + e_{sj} e'_{s'i}) \,.
\end{align}
The expressions above are useful in the evaluation of the squared modulus of transition amplitudes.

\section{One-loop amplitudes}
\label{app:loops}
Below, we list the free electron amplitudes for the four diagrams in Fig.~1 in the non-relativistic limit.

\noindent 1).~{\it Scalar DM with a scalar mediator}.~For the top left diagram in Fig.~1, we obtain the amplitude
\begin{align}
    i\mathcal{M}
    \simeq -i4m_Sm_e
    &\bigg[ 
    \frac{g_2h_1^2}{2}\frac{m_e}{m_S}\frac{\delta_{\lceil h_2 \rceil 0 }}{(4\pi)^2}
        \int_0^{1} \!\!\rd x \int_0^{1-x}\!\!\!\!\!\!\!\!\rd 
        y \frac{x + y + 1}{\Delta} \nonumber\\
    &+
    \frac{g_2h_2^2}{2}\frac{m_e}{m_S}\frac{\delta_{\lceil h_1 \rceil 0 }}{(4\pi)^2}
        \int_0^{1} \!\!\rd x \int_0^{1-x}\!\!\!\!\!\!\!\!\rd y \frac{x + y - 1}{\Delta}
    \bigg] \NRop{1} \,,
\end{align}
where we assume that the only coupling constants different from zero are either $g_2$ and $h_1$, or $g_2$ and $h_1$, while
\begin{align}
\Delta = (x^2 + y^2)m_e^2 + (x + y)(m_M^2 - m_e^2) + xy|\vec q\,|^2 + (1-x-y+2xy)m_e^2\,,
\label{eq:Delta}
\end{align}
with $m_M=m_\phi$.

\noindent 2) {\it Scalar DM with a vector mediator}.~The amplitude for the top right diagram in Fig.~1 is equal to
\begin{align}
    i\mathcal{M}
    \simeq i4m_Sm_e
    &\bigg[
    g_3h_3^2\frac{m_e}{m_S}\frac{\delta_{\lceil h_4 \rceil 0 }}{(4\pi)^2}
        \int_0^{1}\!\! \rd x \int_0^{1-x}\!\!\!\!\!\!\!\!\rd y \frac{x + y - 2}{\Delta}\nonumber\\
    & +
    g_3h_4^2\frac{m_e}{m_S}\frac{\delta_{\lceil h_3 \rceil 0 }}{(4\pi)^2}
    \int_0^{1}\!\! \rd x \int_0^{1-x}\!\!\!\!\!\!\!\!\rd y \frac{x + y + 2}{\Delta}
    \bigg]\NRop{1}
\end{align}
where $\Delta$ is given in Eq.~(\ref{eq:Delta}) with $m_M=m_G$.~In this latter equation, we assume that the coupling constants $g_3$ and $h_3$, or $g_3$ and $h_4$, are the only ones being different from zero in the Lagrangian.\\
\noindent 3) {\it Vector DM with a scalar mediator}.~For the bottom left diagram in Fig.~1, we find the amplitude
\begin{align}
    i\mathcal{M} \simeq -i4m_Xm_e
    &\bigg[  
    \frac{b_2h_1^2}{2}\frac{m_e}{m_X}\frac{\delta_{\lceil h_2 \rceil 0 }}{(4\pi)^2}
        \int_0^{1}\!\! \rd x \int_0^{1-x}\!\!\!\!\!\!\!\!\rd y \frac{x + y + 1}{\Delta} \nonumber\\
    & +
    \frac{b_2h_2^2}{2}\frac{m_e}{m_X}\frac{\delta_{\lceil h_1 \rceil 0 }}{(4\pi)^2}
        \int_0^{1}\!\! \rd x \int_0^{1-x}\!\!\!\!\!\!\!\!\rd y \frac{x + y - 1}{\Delta}
    \bigg] \NRop{1} \,,
\end{align}
where $m_M=m_\phi$ in $\Delta$, and either $b_2$ and $h_1$, or $b_2$ and $h_1$ are the only non-zero coupling constants.

\noindent 4) {\it Vector DM with a vector mediator}.~Finally, the amplitude for the bottom right diagram in Fig.~1 is 
\begin{align}
    i\mathcal{M} \equiv i4m_Xm_e
    &\bigg[
      b_3h_3^2\frac{m_e}{m_X}\frac{\delta_{\lceil h_4 \rceil 0 }\delta_{\lceil b_4 \rceil 0 }}{(4\pi)^2}
            \int_0^{1}\!\! \rd x \int_0^{1-x}\!\!\!\!\!\!\!\!\rd y \frac{x + y - 2}{\Delta}\nonumber\\
    & +
        b_3h_4^2\frac{m_e}{m_X}\frac{\delta_{\lceil h_3 \rceil 0 }\delta_{\lceil b_4 \rceil 0 }}{(4\pi)^2}
            \int_0^{1}\!\! \rd x \int_0^{1-x}\!\!\!\!\!\!\!\!\rd y \frac{x + y + 2}{\Delta}\nonumber\\
    & +
        \frac{b_4h_3^2}{2}\frac{m_e}{m_X}\frac{\delta_{\lceil h_4 \rceil 0 }\delta_{\lceil b_3 \rceil 0 }}{(4\pi)^2}
            \int_0^{1}\!\! \rd x \int_0^{1-x}\!\!\!\!\!\!\!\!\rd y \frac{x+y-1}{\Delta} \nonumber\\
    & +
        \frac{b_4h_4^2}{2}\frac{m_e}{m_X}\frac{\delta_{\lceil h_3 \rceil 0 }\delta_{\lceil b_3 \rceil 0 }}{(4\pi)^2}
        \int_0^{1}\!\! \rd x \int_0^{1-x}\!\!\!\!\!\!\!\!\rd y \frac{x+y+1}{\Delta} 
    \bigg]\NRop{1} \,,
\end{align}
where only $b_3$ and $h_3$, $b_3$ and $h_4$, $b_4$ and $h_3$, or $b_4$ and $h_4$ are simultaneously different from zero, and $m_M=m_G$.~Inspection of the above equations shows that all non-relativistic amplitudes listed here can be written in terms of a single integral function of the integer number $a$, namely

\begin{align}
    I(a) = \int_0^1\!\!\!\!\rd x\int_0^{1-x}\!\!\!\!\!\!\!\!\rd y\frac{x+y+a}{\Delta} \,.
    \label{eq:I}
\end{align}

\providecommand{\href}[2]{#2}\begingroup\raggedright\endgroup

\end{document}